\documentclass{article}

\usepackage{arxiv}

\usepackage[utf8]{inputenc} 
\usepackage[T1]{fontenc}    
\usepackage{hyperref}       
\usepackage{url}            
\usepackage{booktabs}       
\usepackage{amsfonts}       
\usepackage{nicefrac}       
\usepackage{microtype}      
\usepackage{lipsum}
\usepackage{graphicx}
\usepackage{biblatex}
\title{Characterizing Certain DNS DDoS Attacks}

\author{
Ren\'ee Burton \\
Cyber Intelligence \\
Infoblox \\ 
\texttt{rburton@infoblox.com} \\
  }

\begin{document}
\maketitle

\begin{abstract}
This paper details data science research in the area of Cyber Threat Intelligence applied to a specific type of Distributed Denial of Service (DDoS) attack. We study a DDoS technique prevalent in the Domain Name System (DNS) for which little malware have been recovered. Using data from a globally distributed set of a passive collectors (pDNS), we create a statistical classifier to identify these attacks and then use unsupervised learning to investigate the attack events and the malware that generates them. The first known major study of this technique, we discovered that current attacks have little resemblance to published descriptions and identify several previously unpublished features of the attacks. Through a combination of text and time series features, we are able to characterize the dominant malware and demonstrate that the number of global-scale attack systems is relatively small. 
\end{abstract}

\section{Introduction}

In the field of Cyber Security, there is a rich history of characterizing malicious actors, their mechanisms and victims, through the analysis of recovered \emph{malware}\footnote{We use the term malware here to mean the software used for malicious means, not limited to such software on compromised devices.} and related event data. Reverse engineers and threat analysts take advantage of the fact that malware developers often unwittingly leave fingerprints in their code through the choice of libraries, variables, infection mechanisms, and other observables. In some cases, the Cyber Security Industry is able to correlate seemingly disparate pieces of information together and attribute malware to specific malicious actors. Identifying a specific technique or actor involved in an attack, allows the Industry to better understand the magnitude of the threat and protect against it. The study of cyber actors, their motives and mechanisms, falls within Cyber Threat Intelligence, an area traditionally dominated by security analysts utilizing deep domain expertise, but increasingly open to data science methodologies. The malware itself is critical to the vast majority of work of this type.   

In this paper, we address a class of Distributed Denial of Service (DDoS) attacks for which malware is rarely obtained, but which remains a persistent presence on the Internet. Motivated by the need to help threat analysts more easily identify and characterize threats, the results of this research demonstrate a dramatic change in these attacks in recent years, made visible through statistical analysis and unsupervised machine learning. Leveraging over six months of real data, we identify a number of features that allow us to cluster the majority of attacks into a handful of families and in some cases completely reverse the algorithm for generating the attack. Data analysis shows that actors have changed their modus operandi, making detection and correction more difficult for Internet providers. We developed a highly accurate statistical classifier to identify the attacks and engineered features not previously mentioned in the literature, some of which may apply to other cyber problems. 

 The attack, known as a Slow Drip, Random Subdomain, and Water Torture Attack, among other names, targets the Domain Name System (DNS), a globally distributed database that provides critical functionality for the Internet. First observed in 2009~\cite{chinaAttack}, the attacks emerged with tremendous strength in early 2014. While their scale has varied tremendously, they remain a daily fixture of the cyber landscape today. The DNS community has focused primarily on mitigation of ongoing attacks, largely through scaling the ability of systems to handle additional load. Research that attempted to address detection based on attack characteristics was hindered by the limited reporting available or focused on single systems.  This large-scale study, utilizing current, real data, is the first of its kind.
 
 We sought to understand the attack technique, and related malware, as it appears on the Internet today. Specifically, we looked to address common threat intelligence questions:
\begin{itemize}
    \item how many such malware systems exist?
    \item how can we attribute two attacks to the same malware?
    \item has the threat landscape evolved over time, and how?
    \item can we characterize the attack generation algorithms?
\end{itemize}
  
Throughout the remainder of this Section, we introduce DNS and the Slow Drip attack technique, discuss prior work, and provide an overview of our data and methodologies. In Section \ref{sec:detector} we review our method for detecting attacks in passive data and provide a cursory analysis of the attack landscape over a six month period. Section \ref{sec:agenda} discusses feature engineering and includes details of a number of novel features of the attacks. Those features are used to cluster attacks in Section \ref{sec:clusters}, demonstrating that the features are both meaningful and stand over time. The results are examined in the context of our goals in Section \ref{sec:conclusions}. Finally, we discuss ideas for future research. 

\subsection{The Domain Name System (DNS)} \label{sec:dns}

The Domain Name System (DNS) is a global, hierarchical, distributed database which serves, among other things, to map domain names to  Internet Protocol (IP) addresses. While relatively straight forward in concept, in practice the global DNS is complex to the point of being arcane \cite{dnsCamel}. For the purpose of this paper, we introduce a limited scope and vocabulary; the interested reader can find more depth in the operation and security of DNS in \cite{cricket} and \cite{dnsSecurity}. 

The domain name system operates as a query-response protocol, in which a query for a \textbf{fully qualified domain name (FQDN)} is made by a client, or \emph{endpoint}, and is answered via an iterative process known as \emph{resolution}. An FQDN is made up of a series of text \textbf{labels} separated by periods. For example, the FQDN \texttt{www.google.com} has three labels ['\texttt{www}', '\texttt{google}', '\texttt{com}'].  

From a hierarchical perspective, the right-most label in an FQDN is the \emph{top-level-domain (TLD)} and the each subsequent label represents a \textbf{subdomain} of the FQDN created from all the prior labels. For instance \texttt{www.google.com} is a subdomain of \texttt{google.com}, which in turn is a subdomain of \texttt{com}. The term \textbf{domain} is often used to refer to both an FQDN and the scope of its possible subdomains. Thus \texttt{www.google.com}, \texttt{mail.google.com}, \texttt{inbox.google.com}, and \texttt{photo.google.com} are all subdomains of the domain \texttt{google.com}. The TLD is a single label and is considered \emph{public} in the sense that its subdomains are available for registration and not controlled by the TLD. The TLDs are limited in number and controlled by the Internet Corporation for Assigned Names and Numbers (ICANN). In some cases, subdomains of a TLD are also managed as public domains, most notably domains like \texttt{co.uk}, and are considered \emph{public suffixes} and create extended TLDs (eTLD).\footnote{A list of public suffixes is maintained by Mozilla and found at publicsuffix.org} A \textbf{second-level-domain (SLD)} is privately owned and the direct subdomain of a public suffix, or eTLD. Examples of second-level-domains include \texttt{google.com} and \texttt{google.co.uk}. The SLD is sometimes referred to as the base domain. 

Within the Domain Name System, \emph{authoritative name servers} are servers which hold authoritative, or definitive, answers for a certain portion of the database, generally a specific domain. If these authoritative servers are not functioning properly, Internet traffic to the domains for which they are authoritative may be interrupted or completely disrupted. For this reason, companies often have multiple authoritative name servers for their domains.  

While it is possible for an endpoint to resolve DNS queries themselves, in practice, most devices rely on large \textbf{recursive resolvers} to perform resolution on their behalf. Internet Service Providers provide recursive resolvers for their customers, for example. Many recursive resolvers are configured to answer queries only for devices in their network, thereby limiting the resource demands on those network appliances. There are \emph{public resolvers}, such as those operated by Google, openDNS, and Cloudflare, designed to handle recursion for any endpoint selecting their service.\footnote{Endpoints are typically configured with at least one recursive resolver to which they forward their queries.} On the other hand, there are also a large number of devices on the Internet that, through misconfiguration or otherwise, will act as recursive resolvers for any client but are not announced as public resolvers. These are known as \textbf{open resolvers} and are frequently leveraged by cyber actors to anonymize and amplify DDoS traffic.

In another effort to ensure efficiency, the Domain Name System employs \emph{caching} of records at the recursive resolvers. Caching of records prevents unnecessary Internet traffic by caching the records for popular domain names, such as \texttt{www.google.com}. For security purposes, many recursive resolvers also employ negative caching, in which they remember, for some short period of time, that a given FQDN has no answer in the DNS. Queries for non-existent domains typically return what is referred to as an NXDOMAIN response which triggers this form of caching.

\subsection{Slow Drip DDoS Attacks}

Given the fundamental role the DNS plays in the functioning of the Internet, attackers are known to abuse it. In the technique considered in this paper, malicious actors generate a massive number of queries for non-existent domains. Through the resolution process, these queries are forwarded to the authoritative name servers which may become overwhelmed with the unexpected volume. The attack generates non-existent subdomains of a common SLD, with little to no repetition in the queries. This serves to counter caching at recursive resolvers and force as much traffic as possible to the authoritative name servers. All of the queries within an attack will be subdomains of a common SLD, which we refer to as the \textbf{attack domain}. Authoritative servers may attempt to mitigate an attack in a number of ways, including dropping requests from recursive resolvers that transmit too many requests.\footnote{This is known as Response Rate Limiting (RRL).} In a counter play, attackers may utilize open resolvers to diffuse their traffic over a large IP space, maximizing transmission paths and reducing the likelihood of blocked traffic. Some attackers utilize spoofed\footnote{IP Spoofing is the creation of Internet traffic with fake IP addresses.} IP addresses in the request packets, which help both anonymize and diffuse traffic, hindering mitigation.

This attack technique was first recorded in 2009 in China~\cite{chinaAttack}, and again in a large attack against AFNIC in October 2013~\cite{afnicAttack}. Beyond these two attacks, there is no indication the technique was actively used until February 2014 when it became a daily phenomena that disrupted the global DNS and had particularly damaging collateral damage on Internet Service Provider (ISP) infrastructure \cite{nominum1}. During the early years, the attack used queries for pseudo-random subdomains of primarily Chinese-owned domains. The terms Random Subdomain attack and Slow Drip, or Water Torture, attack were derived from these characteristics. While targeted at authoritative name servers, these attacks are often more damaging to the Internet's middle infrastructure~\cite{qtnet}, including recursive resolvers within ISPs, as they were not provisioned for unusually high traffic volumes. Solutions like response-rate-limiting were designed for use at the authoritative server, and are less effective at recursive resolvers where the distribution of client IPs may not trigger such protections even if they were implemented. For practical reasons, most previous analysis of these attacks focused on mitigation. 

\begin{figure}
    \centering
    \includegraphics[width=\linewidth]{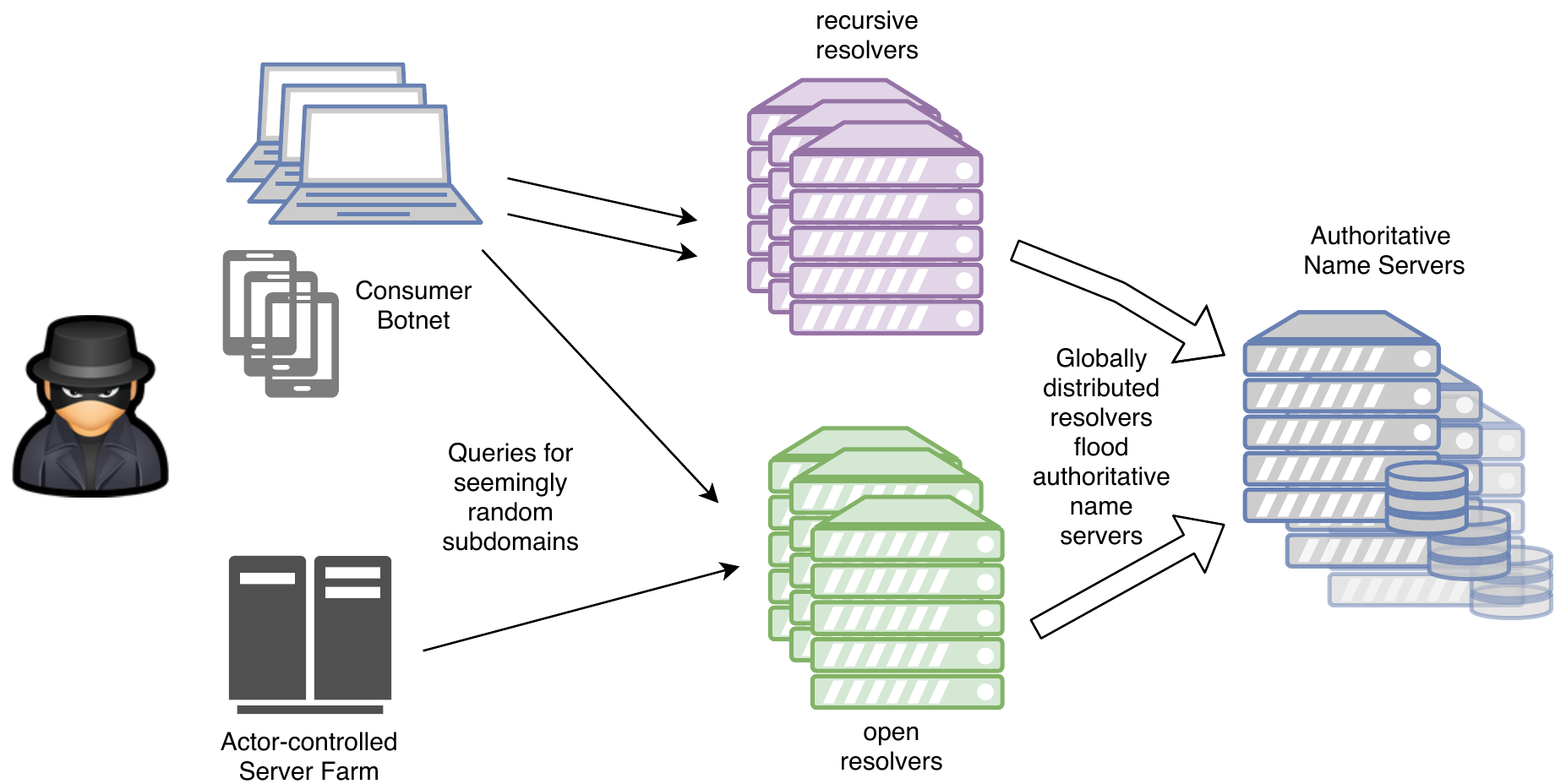}
    \caption{Slow Drip Attack Mechanisms}
    \label{fig:slowdrip}
\end{figure}

While the common domain for an attack is considered the attack domain, the nature of the attack creates a resource exhaustion of the authoritative name servers, not the attack domain or its Internet resources, such as web pages. Beyond potentially disabling the authoritative name server from answering legitimate queries, e.g, for their website, the attacked domain suffers no direct denial of service traffic from these particular attacks. From this perspective, the attack domain itself does not play a significant role in the attack, only the authoritative name server. It is also possible for the attack domain to be operated by the attacker as a mechanism to target a particular name server.

If a name server's resources become exhausted, it may be unable to serve any of its customers, regardless of the requested domain. In reality, many different types of cyber attacks may occur simultaneously, including ones that target the web servers of the attack domain, for example. In the famous Mirai attacks of 2016, a wide range of attacks were deployed simultaneously, and in series, against Dyn authoritative name servers, effectively crippling a large number of popular websites~\cite{mirai}. One of the types of attacks used was Slow Drip. In other cases, malware has been discovered that contains different DDoS mechanisms for domains known to be attacked via Slow Drip; it seems plausible that these attacks were used in conjunction with one another \cite{magicFerret}.

More precisely, an attack consists of queries for a very large number of non-existent FQDNs within a common \textbf{attack domain}. Queries that are part of the attack will have one or more subdomain labels. If we consider the attack domain as an SLD, \emph{attack\_domain}, then queries that are part of the attack traffic will have the form:
\begin{equation}\label{eq:slowdrip}
label_n.label_{n-1}...label_1.attack\_domain
\end{equation}
The string $label_n$ is referred to as the \textbf{prefix}. The attack domain is an SLD, most often containing two labels. The \textbf{suffix} of an FQDN is the parent domain of that FQDN, determined by removing the prefix, and thus of the form $label_{n-1}...label_1.attack\_domain$. Slow Drip attacks are described in the literature as a set of pseudo-random prefixes of a single suffix~\cite{ExploderBot}~\cite{cloudmark}. Previous published examples of confirmed attack domains include \texttt{111f.com}~\cite{ExploderBot} and \texttt{www.dafa888.com}~\cite{afnicAttack}, each of which contained a single pseudo-random label. 

\subsection{Prior Work}

The vast majority of literature on this specific type of DDoS attack has focused on detection and mitigation. This includes methods of determining an ongoing attack from the rate of queries~\cite{heavyHitters} and the randomness of the FQDNs \cite{YuyaTakeuchi2016}. In \cite{heavyHitters} the authors propose an algorithm for identifying ongoing attacks within a recursive resolver or name server which leverages a fixed size cache to identify domains that are seen frequently and, for the case of Slow Drip attacks, suddenly frequently. Their solution for inline detection promises to require less resources and afford more timely identification than traditional approaches. In \cite{YuyaTakeuchi2016}, they propose a means to detect attacks based on the queried FQDN alone and introduce a set of text-based features for this purpose. Unfortunately, this approach made assumptions about the randomness in the traffic not found in real data and therefore is of limited use operationally. 

Only a small percentage of the literature contains analysis of the attack mechanisms and their victims, and none has undertaken a global study over a long period of time. An early expository~\cite{Joost} studied several months of traffic from a personal network in 2014. The author found distinct signatures in the IP packets of an attack and demonstrated large-scale IP spoofing.\footnote{The primary source for this analysis was ICMP error messages, which allowed the author to make inferences about the attack over time.} They augmented statistics about the attacks with some investigation of the attack domains, showing that the majority of these were Chinese owned.  Patterns within the characters of attacks were described in blogpost~\cite{cloudmark} in 2014.\footnote{We are able to verify that the attacks discussed in the blog are all related to the actor discussed in~\cite{ExploderBot}.} 

In work that considered a several DDoS techniques, Qihoo360~\cite{Qihoo360} used unsupervised learning to study traffic observed at honeypot\footnote{A honeypot is a server set up as a decoy to lure cyber actors and to detect, deflect or study cyber attacks.} sensors in order to understand "who is being attacked by what botnet families under which c2 controllers with what set of attack parameters". They began tracking DDoS botnet families in 2014 and covered different forms of DDoS attacks, including two that are DNS-based. Their approach used clustering on packet-level features. One of these clusters, which they call $dns\_cls1$ is a random subdomain attack. The examples they provide are consistent with those found by \cite{Joost} and confirmed by \cite{ExploderBot} as part of a single attack system.\footnote{In \cite{Qihoo360}, they also claim that these prefixes are consistent with the Elknot/BillGates malware.}

In \cite{ExploderBot}, a single DDoS attack system is studied in depth. First documented by \cite{Joost}, this attack system, dubbed \emph{ExploderBot} by the authors, was initially seen in February 2014. These attacks were observed almost continuously through mid-2016, after which their activity was interrupted by multiple long periods of silence. The analysis of \cite{ExploderBot} concluded that:
\begin{itemize}
    \item the attacks contained a single pseudo-random alphabetic label of length 1-16 which followed a precise pattern,
    \item the traffic was not generated by a consumer-device botnet, 
    \item the packets utilized very-large scale IP spoofing, 
    \item the attacks did not appear to favor open resolvers to diffuse traffic, and
    \item the attacks were the work of a single actor.
\end{itemize}
The vast majority of domains either had Chinese-operated authoritative names servers or were registered in China~\cite{ExploderBot}~\cite{Joost}. We independently confirmed the pseudo-random attack pattern and the attack tempo seen in~\cite{ExploderBot}, \cite{Joost}, and \cite{cloudmark} using historic data from 2016-2017.

In the vast majority of reporting on Slow Drip DDoS attacks, the examples cited to demonstrate the attacks can be tied back to the ExploderBot system. The Mirai malware is also known to contain this technique, as documented in \cite{mirai}, and uses a twelve-character alphanumeric label. However, the Mirai malware does not transit the DNS normally, instead sending traffic directly to IP addresses configured in the malware, and we have not observed this traffic in our data. Beyond these two systems, little is known about the malware and actors that generate these attacks. 

\section{Data and Methodologies}

By early 2018, threat analysts noted that it was increasingly difficult to determine whether high volume traffic was part of a Slow Drip attack or due to some other errant Internet phenomena.\footnote{Personal conversations between this author and professional cyber threat analysts.} One analyst provided examples of spikes in AirBnB-related domains that appeared suspicious in terms of their volume and variety, but in which many of the subdomains could be interpreted as locations in China. In other examples, the subdomains contained a hodgepodge of unusual, but not random, words. Another analyst highlighted attacks that leveraged domains known to be associated to the hacker group MageCart. They wondered whether MageCart was branching from credit card theft into the area of DDoS, or whether some other actor was attacking these domains. Were these attacks related to those seen on the AirBnB domains, they asked, and were they Slow Drip attacks at all? The extraordinarily large volume and variety of data makes human analysis time consuming and faulty.

Motivated to reduce the burden on threat analysts in identifying and characterizing the threat of these attacks, we set out to understand what Slow Drip systems were active and how their techniques, tactics, and protocols (TTP) differed from those of ExploderBot and Mirai. 

\subsection{Data Sources}

This research relied on two independent accesses to \emph{passive DNS} (pDNS) records, obtained by recording DNS query-response events at recursive resolvers and including over twelve billion passively observed records per day. The majority of the data was collected between large recursive resolvers and authoritative name servers. A smaller source included data from several open resolvers.\footnote{These are recursive resolvers which will resolve queries for any client but are not publicly advertised.} The data sources included recursive resolvers present on four continents. The smaller open resolver sources collected approximately 10 million records daily. The data used was obtained over seven months, from June through November 2018, and again in the month of January 2019.

In order to build a data set for analysis, we developed a statistical classifier, described in Section \ref{sec:detector}, to identify attacks within normal traffic. Attacks are detected using \emph{events} comprised of query-response pairs, aggregated by the SLD. For each detected attack, we extract the following fields from all events related to the domain for that day, regardless of their resolution:
\begin{itemize}
    \item UTC timestamp of observation, 
    \item fully qualified domain name (FQDN, qname),
    \item query type (qtype), 
    \item response code (rcode)
\end{itemize}
While \cite{ExploderBot} and \cite{Joost} used other elements of the IP and UDP header, such as the time-to-live and port fields, these were not available to us.\footnote{UDP is the User Datagram Protocol used most often to encode DNS queries for transmission.}
  
\subsection{Collection Bias and Noise}

\emph{Collection bias}, also called observation bias, is a potential source of problems for this type of research. Without access to the actual malware, we infer information based on observable data, which is inherently limited. While our data collectors are globally distributed and cover a range of network types, it is possible that large scale attacks still do not pass our sensors, or occur at volumes too low for our detector threshold. As we attempt to study attacks over time, our collection is also susceptible to changes in the underlying traffic due to variability in Internet routing, among other reasons. 

We also have to contend with a wide range of \emph{noise} in the data. DNS traffic is naturally prone to noisy data, as queries are passed through the resolution process largely unchecked until they reach the authoritative server. The multitude of applications running on endpoints can create illegitimate queries.\footnote{Google Chrome is known to create unresolvable pseudo-random queries, for example.} Moreover, users can  accidentally drive traffic through the DNS.\footnote{For example, a user might mistype a website name in their browser.} In addition, while we expect a single attack query to create a single event in the data, recursive resolvers and other devices in the path of the query may replicate the query or even initiate new queries. In some portions of this analysis, successful resolution of a domain is considered noise in the sense that it further skews the data and makes analysis more difficult. The impact of noise on our analysis is largely dependent on the popularity of the domain, as more popular domains are more likely to produce noisy data. The problem of noise was addressed through normalization and considering only non-resolved queries.    

\subsection{Methodologies}

Attack data was obtained through the creation of a statistical classifier, as described in Section~\ref{sec:detector}, that identified anomalies in traffic within and across days. The classifier was tested on previously published attacks that occurred between January 1st, 2017 and May 31st, 2018 to ensure it's reliability and the results of newly identified attacks manually analyzed. From the resulting detection data, events were summarized into a small set of fields for the purpose of analysis and feature engineering. 

As detailed in Section~\ref{sec:agenda}, exploratory analysis of the resulting attack records led to a number of strong features that segregated attacks into several distinct groups. While a wide array of potential features were analyzed for their ability to distinguish attack algorithms from one another, the most promising were derived from lexical and time series analysis, along with the variety of query types used during an attack. Anomalies in the attack traffic were discovered by measuring the attack events against a baseline of global DNS traffic over time. Character distributions proved to be particularly useful in separating attacks. To take advantage of this, archetypal attacks were identified through unsupervised machine learning and then used to create a distance measure used in more comprehensive clustering later. 

Features were derived using three months of data and clustered as described in Section~\ref{sec:clusters}. Finally, to address the potential of transient features derived from the Summer 2018 data, the same features and cluster analysis was applied to January 2019 data and the results compared for consistency.

\section{Attack Detection}
\label{sec:detector}

Large-scale Slow Drip attacks are easily detected at the authoritative name server and can sometimes be detected in live traffic at recursive resolvers with methods like those described in \cite{heavyHitters}. However, our research is based on passive DNS analysis. Detecting attacks in passive DNS collection is made difficult by the scale and variety of global DNS traffic, as well as the nature of the attack. Traffic created by anti-virus products and content delivery networks (CDNs) has patterns that resemble a number of cyber attacks, including Slow Drip DDoS attacks. Attempts to identify attacks using simple measures on the FQDN, as in \cite{YuyaTakeuchi2016} will fail in this environment, creating a large number of false positives, as will approaches based on measurements of traffic volumes within a day. To overcome these issues, we use both current and historical data in the detection algorithm. 

The detector is a statistical classifier that identifies potential attacks within daily traffic. It contains two stages that identify anomalous domain activity within a day of pDNS records which is also anomalous in comparison to the domain's prior history. To trigger the detector a domain (SLD) must have an unusually high number of unresolved queries within a day and, additionally, the number of unique subdomains must be abnormal in comparison to the domain's recent past.\footnote{We have also observed these attacks against so-called wildcard domains, in which every subdomain is resolved, however, we do not consider that variation in this study.} To test performance, the classifier was run on historical data covering ten days in 2017 and 2018 for which Slow Drip attacks were reported in \cite{ExploderBot} and which the openDNS Twitter feed DNSStream contained alerts for Resource Exhaustion Attacks, generally synonymous for Slow Drip attacks in DNS.  The results were manually reviewed for accuracy.

The detector first groups all of the events for a given data source by  domain, or SLD, and counts the number of distinct unresolved queries.\footnote{We consider all unresolved queries, not just NXDOMAIN responses.} In the larger data source, the initial twelve billion records is reduced to approximately 225 million domains. Given the distribution of unresolved query counts across this entire set of domains, we determine the quartiles, $Q1$ and $Q3$, that represent the number of events below which the bottom 25\% of domains, and above which the top 25\%, lie, respectively. The Inter-Quartile Range (IQR) is defined as $Q3 - Q1$, and \emph{outliers} are those domains, $d$, for which the number of unresolved queries, $N(d)$, satisfies
\begin{equation}
    \label{eq:outlier}
    N(d) > 1.5*IQR + Q3.
\end{equation}
This formulation identifies a set of domains which have an unusually high number of unresolved queries, with respect to all other domains on a given day. As concrete examples, on August 1st, 2018, the detector found unusually high query volumes for \texttt{uberinternal.com} and \texttt{icelandairlabs.com}. When compared to previous day's volumes, \texttt{uberinternal.com} had a 5600-fold increase in activity and \texttt{icelandairlabs.com} showed no queries on the prior day.  

Unfortunately, determining statistical outliers in this way is not sufficient for identifying Slow Drip attacks in passive DNS collection. Used alone, this will lead a number of false positives, particularly resulting from anti-virus product and content delivery networks. To address this issue, we also compare the traffic with the recent past. Attacks often cross a date boundary, and detection will be missed if the comparison is only made with the previous day. On the other hand, it is important to compare recent activity to ensure that the large number of queries represents a true change. For this reason, we use a two day separation for detection. The second stage of the detector looks for a dramatic increase in the number of subdomains observed for domains raised during the first stage. This is accomplished by computing the change in the number of unique subdomains between the two dates for all second-level-domains, calculating the quartiles for the resulting distribution, and determining the outlier threshold analogous to that in Equation~\ref{eq:outlier}. Domains raised in the first stage, which are also anomalous in the second stage, are considered likely Slow Drip attacks. As a concrete example, \texttt{uberinternal.com} showed a 214k-fold increase in the number of subdomains queried between days.

These two stages create thresholds above which unresolved queries for a domain are considered a Slow Drip attack. The thresholds for an attack vary for each data source and are dependent on the normal volume observed in that source. The combination of anomaly detection with historical perspective serves to weed out false positives. In evaluating nearly two thousand attacks over a six month period, across multiple data sources, no false positives were found. On the other hand, this approach can have false negatives, that is, it might miss attacks for which we do not observe enough traffic.

In mid-January 2019, the amount of Slow Drip traffic increased dramatically over the previous year. While having the same characteristics as Slow Drip DDoS attacks, the volume of the activity was much lower than necessary for an effective denial of service. The domain characteristics were quite similar, however, to others studied in this research.  This activity was noted by top-level-domain administrators, as in~\cite{klaus}.\footnote{In a follow on email discussion, the author noted that they had observed this activity at lower levels for approximately one year, and found it hard to detect.} Unlike large-scale attacks, this activity is much harder to detect at the authoritative name servers. It is, however, consistently detected by the algorithm discussed here.

\section{Attack Landscape}

To understand the overall attack landscape, meaning the tempo and general characteristics of Slow Drip attacks, we analyzed attacks discovered with the detector described in Section~\ref{sec:detector} over a six month period (June - November 2018), where an \textbf{attack} was identified by a domain and a date. During this time, there were 1949 separate attacks detected, covering 1418 domains, with a median of seven attacks daily and eight days with over twenty-five targeted domains. On days with a large number of attacks, the attacks often contained groups of related domains. For example, we found ten attacks on August 12, 2018 targeting domains related to Western Union. On other days, the relationship between domains was not readily apparent. The volume of attacks in the open resolver set is, as expected, much smaller, and with more variance. It includes 221 attacks cover 143 unique domains over the same time period.

We were particularly interested in finding attacks observable in independent data sources, including data from \emph{open resolvers}. The use of open resolvers by attackers makes both detection and source identification of attacks more difficult by distributing the attack.\footnote{These resolvers are not publicly announced as recursive resolvers, but are easily detected by daily scanning of the Internet by organizations such as Censys.} Discovering the same attack across sources lends support to the accuracy of our detector and indicates the likely breadth of an attack. Nearly 100 attacks and a total of 48 domains were observed in multiple data sources that included at least one open resolver. The inclusion in both sets signifies an actor tactic to leverage open resolvers. The domains in the intersection are both unsurprising, such as \texttt{airbnb.com}, and curious, such as a series of domains beginning with \texttt{sc-} seen in June 2018 that appear to have no legitimate purpose or common relationship.  
 
In spite of a very strong signature for ExploderBot Slow Drip traffic, we found no attacks of this kind during the evaluation period. We detect this actor consistently through historical data and so while it remains possible that our data sources are simply not observing the attacks, it appears more plausible that the actor has remained quiet through the latter half of 2018.

This research focused on the malware, specifically the generation of subdomains used in the attacks, and we did not do an in-depth analysis of the victims. We did find a prevalence of high profile domains, such as those owned by AirBnB, Sony, DotDash, and Toyota present in the attacks. In the final clusters, described in Section~\ref{sec:clusters}, we see groups of well-known domains such as these, as well as clusters containing Pharmaceutical and Banking Industry domains. At the same time, almost 38\% of the domains were not in the top one million most commonly observed domains within our datasets and many showed little sign of normal traffic. Some examples of these domains include \texttt{nspk.ru} and \texttt{verimi.de}. In these cases, the motivation for a DDoS attack is particularly unclear. In contrast to observations of Slow Drip attacks in 2014-2015, we did not find a preponderance of Chinese domains and name servers.

\section{Attack Features}\label{sec:agenda}

Slow Drip attack traffic is most often described as random subdomains of an attack domain, that is, queries for non-existent FQDNs of the form
\begin{equation}
\label{eq:randomPrefix}
    random.attack\_domain,
\end{equation}
where $random$ is a single pseudo-random label and $attack\_domain$ is an SLD. In some cases, as described in~\cite{cloudmark}, a second label was observed, leaving queries of the form
\begin{equation}
\label{eq:withSuffix}
    random.fixed\_label.attack\_domain
\end{equation}
The random label, or \textbf{prefix}, is generated by a portion of the malware we call the \textbf{attack generator}. Without a copy of the malware, it might be impossible to fully determine how specific attack traffic is generated. Given that attack generators are software written by individuals, the hope is that the traffic carries the fingerprints of the designer and is distinct enough that it can be correlated across attacks.\footnote{Two different pieces of malware may share the same generator, particularly if it is simplistic, and two different cyber actors may utilize the same malware for different attacks.}

A \textbf{feature} in machine learning is a measurable characteristic of the data. In \textbf{clustering}, we want to identify features that are useful for grouping subsets of data and for isolating groups from one another. This is a \textbf{unsupervised learning} approach, as we don't know in advance the true nature of the groups, or clusters, and the algorithm learns these from the data itself. A feature in the data is considered a \emph{strong feature} if it can reliably separate groups of data, and a \emph{weak feature} if it demonstrates a measurable characteristic for separating data, but can't be used alone to so. A number of weak features might be combined to reliably separate data. These terms are subjective, but useful in understanding how features help in the machine learning process. \textbf{Feature Engineering} is the iterative process of identifying, evaluating, and selecting features. Successful use of machine learning to solve problems is highly reliant on proper feature engineering, much more so than the choice of the machine learning algorithm itself. Feature engineering also a laborious process, and researchers often rely on previous work to obtain features for their experiments. 

In the case of Slow Drip attacks, little insight into the attack features exist in the literature and that which does was specific to certain systems. As a result, we undertook the feature engineering process from scratch. In doing so, we sought to find features of the data that would allow us to group attacks that were likely created by the same algorithm, that is, the process for creating the attack queries, and that could separate these from unrelated attacks. This process begins with exploratory data analysis (EDA), in which large random samples of the attack data are analyzed in-depth. After initial exploratory data analysis, we validated and refined the feature set using the event records from 435 attacks identified between June and August 2018. These features were then applied to cluster attacks into related groups.  

\subsection{Exploratory Data Analysis}\label{sec:eda}

During the initial review of the data, it was readily apparent that the structure of these attacks was quite different than that previously reported, as described in Equations~\ref{eq:randomPrefix} and \ref{eq:withSuffix}. In this section we provide a brief overview of the results of Exploratory Data Analsysis (EDA) performed on a sample of attacks, and then dive into the resulting features more deeply in the sections that follow. The Appendix includes further visualizations and detailed discussion of the feature engineering process. 

Perhaps most evident in the differences was the use of what appeared to be dictionary terms, rather than, or in addition to, pseudo-random values in the subdomain labels. Additionally, many of the attacks contained FQDNs with several labels, in stark contrast to the single pseudo-random label noted in the literature. Here are a few examples:
\begin{itemize}
    \item \texttt{ent254.sharepoint.hp.com}
    \item \texttt{passyourdrugtest.airbnb.com}
    \item \texttt{ltrefgt.byairbnb.com}
    \item \texttt{rtx.bjbgp.bjbgp.bjbgp.91y.com}
    \item \texttt{godoid-028.prod.ap-northeast-1.int.vidible.tv}
    \item \texttt{www.diskgas.api.csd.bitwala.com}
\end{itemize}
Even this limited set of samples reveals the problems experienced by threat analysts in evaluating suspected Slow Drip traffic. The prefixes can hardly be described as pseudo-random. 

\begin{figure}
    \centering
    \begin{minipage}{0.45\textwidth}
    \centering
    \includegraphics[width=0.95\textwidth]{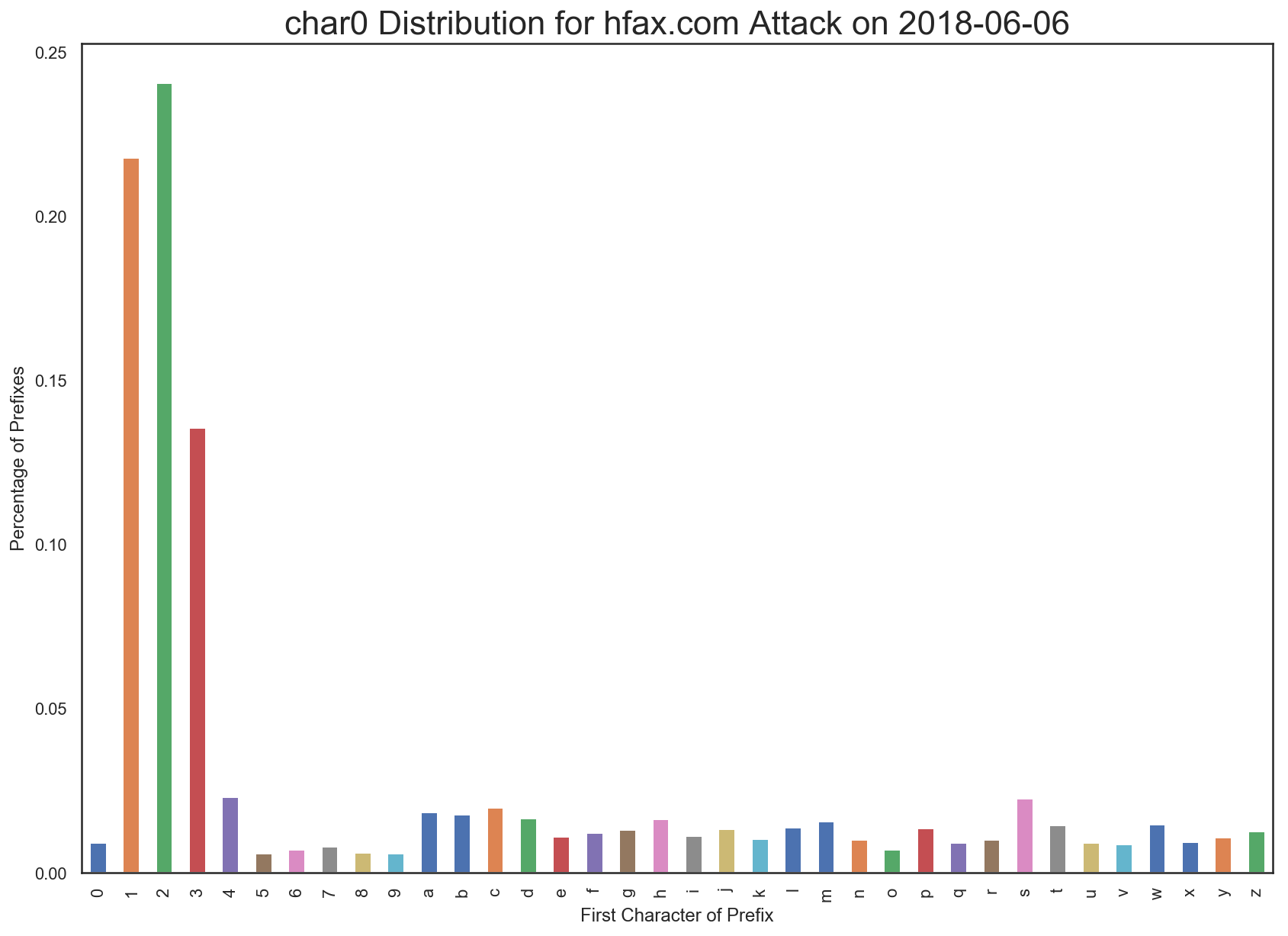}
    \caption{The distribution of first characters seen in an attack on \texttt{hfax.com} shows that prefixes most often started with digits.}
    \label{fig:hfaxDist}
    \end{minipage}\hfill
    \begin{minipage}{0.45\textwidth}
    \centering
    \includegraphics[width=0.95\textwidth]{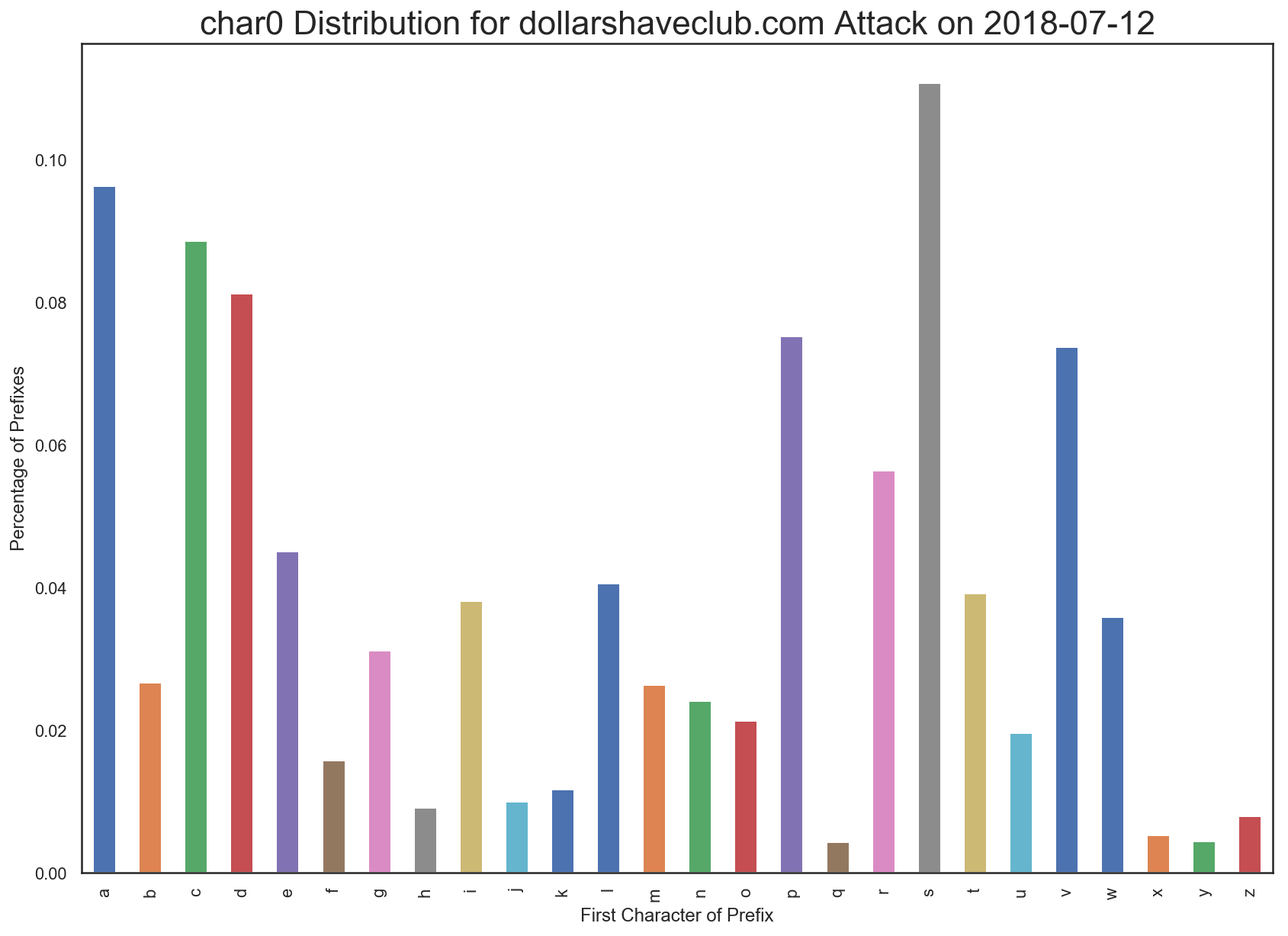}
    \caption{The distribution of first characters seen in an attack on \texttt{dollarshaveclub.com} shows that prefixes started with letters.}
    \label{fig:dollarDist}
    \end{minipage}
\end{figure}

When considered in aggregate, the prefixes in the sample data had strong characteristics that distinguished one attack from another, and were unlike pseudo-randomly generated data. This feature was quantified by creating \textbf{unigram character distributions} for each attack. This is accomplished by converting all of the prefixes observed in an attack to a set of individual characters, called a unigram, and counting the number of times each character is present in the set, creating a distribution. The resulting count distributions can be compared, both visually and statistically, for similarity. As an example, two sample character distributions are shown in Figures \ref{fig:hfaxDist} and \ref{fig:dollarDist}. These distributions, derived from millions of observations, are very clearly distinct.

In the sample data, only a handful of distinct unigram character distributions were present, meaning that attacks used characters in the prefixes proportional to a select few and could be used to separate groups of attacks by the characters used in their prefixes.\footnote{While we used a single character, or unigram, here, it is common in text-based machine learning problems to create features based on longer sub-strings, such as bigrams or trigrams. These were not used because the unigram character distribution provided a strong feature and required less computational resources.} The process of generating unigram character features for use in clustering attacks was a multi-stage process and is detailed in Section~\ref{sec:unigram}.

Also notable was the large number of labels commonly seen in normal DNS traffic, such as \texttt{prod} or \texttt{cloud}. A single attack often included both common terms and a strange assortment of uncommon labels. This exemplified another challenge for threat analysts in reviewing attack queries. In an effort to determine whether these attacks utilized the same generators, we extracted very long labels that were present across multiple attacks. Doing so led to a handful of surprising labels, e.g., \texttt{007dapotianmuweijiaomenduchang}, common to several seemingly independent attacks. This led to the hypothesis that an unusual dictionary was used in the construction of some attacks; the resulting features are described in Section~\ref{sec:dns_enum}.

Time series analysis, that is, the analysis of attack events considered in time order, proved one of the most fruitful approaches to explore features within the sample data. Visualization of various summary statistics of the events illuminated distinguishing characteristics between attacks. As an example, some attacks contained FQDNs with a wide range of prefix lengths and the proportion of each length remained relatively fixed over time, while other attacks contained only prefixes of a fixed length at any given time, as seen in Figure \ref{fig:hfax}. Most prevalent among these characteristics was the lexicographic ordering of the FQDNs, when considered as a series of strings ordered in time, in many attacks. Although deriving a quantifiable measurement of that feature proved challenging, as discussed in Section~\ref{sec:timeseries}.

A number of other statistical features were apparent in the data and differed significantly from that which would occur if the attack traffic was similar to Equation~\ref{eq:randomPrefix}. For each of our findings in the exploratory data analysis phase, we engineered and evaluated features in Spark across a data set including 435 attacks, comprised of over a billion event records. In the sections that follow, we describe the features in depth and discuss their promise for clustering attacks into common attack generators. 

\subsection{Unigram Character Features}\label{sec:unigram}

To take advantage of the similarity, and differences, observed in character distributions between attacks, we designed a process that allowed us to measure the difference between \textbf{unigram character distributions} for the prefixes in each attack against a set of fixed distributions, creating a feature that could be computed reliably over time. Given two distributions, the \textbf{Jensen-Shannon (J-S) distance} \cite{JSD} is a distance measure with a real number value in the interval $[0,1]$, where smaller values indicate that the two distributions are more similar.\footnote{The Jensen-Shannon distance between two probability distributions is the square root of the Jensen-Shannon divergence of the distributions.} To create a feature to be used in the final clustering of attacks, we first used the Jensen-Shannon distance alone to identify a set of \textbf{archetype attacks}. The distance between two attacks is defined as the J-S distance between their respective distributions. For each of the 435 attacks, unigram character distributions were computed and the pairwise Jensen-Shannon (J-S) distance between these distributions calculated. We hypothesized that attacks created by the same malware, or attack generator, would have similar character distributions and therefore small J-S distances. Computing the pairwise distance between all attacks over time, however, is unwieldy, at best and not an effective feature. Instead we clustered the initial attacks using only this distance measures and identified a small set of attacks that represented large clusters. The specific process for this feature engineering is described in detail below. This approach does not attempt to represent all attack generators, but instead provide a statistical feature that measures the distance from a given attack to a handful of other fixed attacks. We have not seen this approach of converting distribution similarities into a distance metric and then utilized as a clustering feature described elsewhere in the literature. This section elaborates on the process used to select archetype attacks.

We considered two different unigram distributions for each attack. The first was the overall unigram character distribution of the set of unique \textit{prefixes}, or $label_n$, over all FQDNs observed in an attack. We refer to this as the \textbf{overall character distribution}. We also considered the distributions of only the first character of each prefix within an attack. We refer to these distributions as \textbf{char0 distributions}. We then consider the underlying dictionary of characters observed across all the attacks. A total of sixty-six distinct characters were present in the labels across all attacks.\footnote{Many of these characters are not valid in DNS, however, the queries will still transit the DNS for resolution.}  

DBSCAN \cite{DBSCAN} was used to cluster the attacks via pairwise distances. This algorithm was chosen for clustering attacks by character distribution because it does not require the number of clusters to be selected apriori and it identifies \emph{outliers}. Each cluster is comprised of \emph{core points} and \emph{boundary points}. The algorithm has two parameters, $eps$, which specifies the distance the algorithm searches for additional cluster members from core points, and $min\_points$, the number of points that must exist within epsilon range for a point in the cluster to be considered a core point. Any point which is further than $eps$ away from all others is an \emph{outlier}. In the application of DBSCAN to the unigram character distributions, the \textbf{distance between two attacks} is the J-S distance between the associated character distributions.

The goal was to identify tight clusters that might represent attacks created by the same attack generator. For this reason, we did not attempt to account for all of our data and were unconcerned about outliers.\footnote{Outliers in clustering algorithms are subjective and a result of thresholds applied to define cluster boundaries. An outlier is any point that lies outside of all cluster boundaries and can be thought of as points that are dissimilar, according to the features used, from all others.} The most central distribution within a cluster became the representative, or \textbf{archetype}. The feature for use in more comprehensive clustering was the distance of each attack from a small set of archetype attacks. 

This approach proved very successful. While the clustering was based on attacks, uniquely identified by a date and domain pairing, the clusters \emph{effectively} represented domains, meaning that a given domain, even if attacked on several days, was usually found in only one cluster. This is consistent with attacks being created by the same generator, or set of generators, for a given domain over time.  In other words, if we observe attacks against a domain like \texttt{uberinternal.com} or \texttt{airbnb.com} over several distinct days, the prefixes generated have similar character distributions.   

\begin{figure}
    \includegraphics[width=\linewidth]{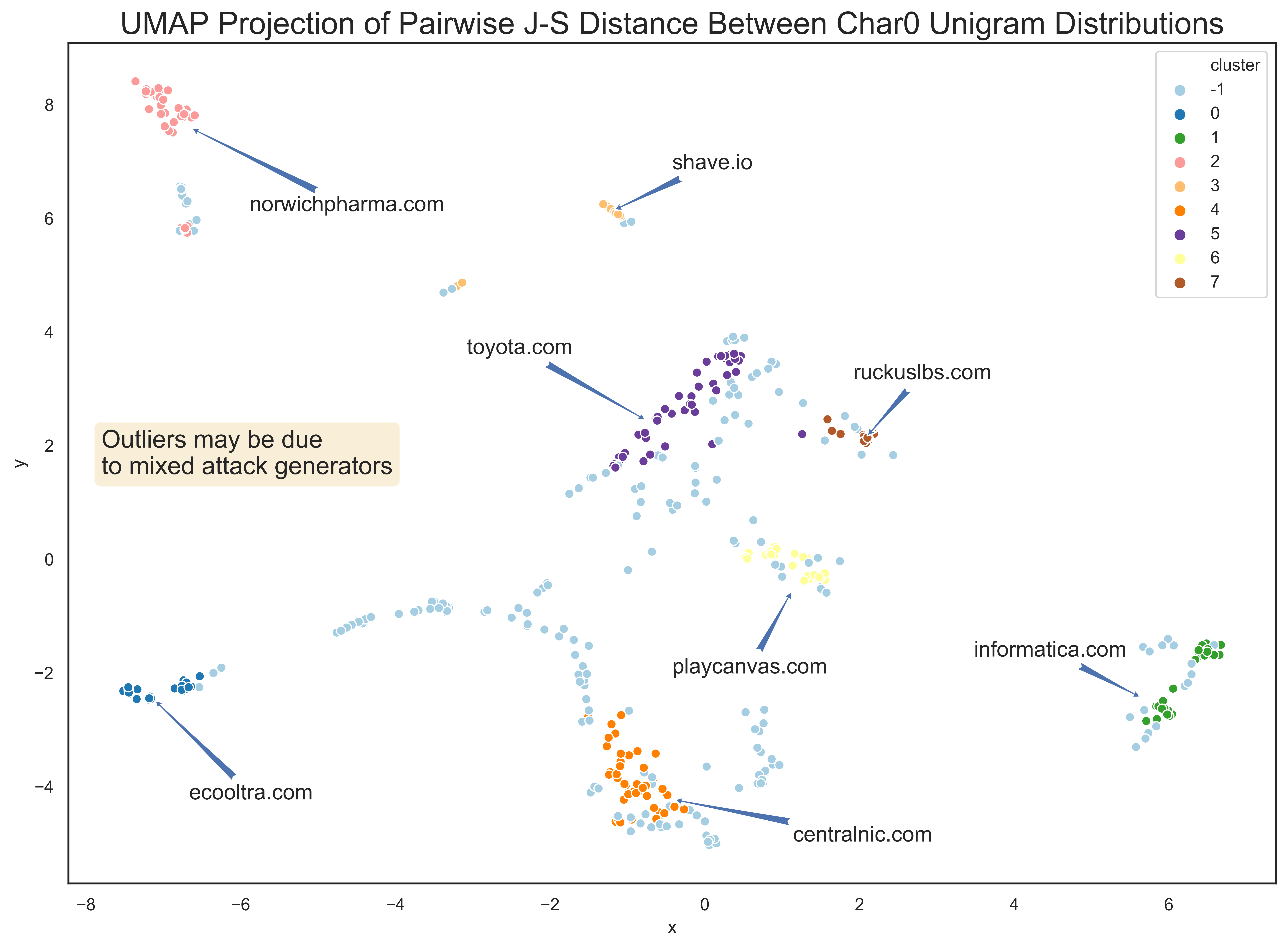}
    \caption{Clusters of attacks based on the J-S Distance between the char0 distributions.}
    \label{fig:umap_char0}
\end{figure}

The overall unigram distributions led to four clusters, and using the char0 distributions we found eight clusters containing more than ten attacks. A visualization of the attacks, clusters, and the points that weren't labeled, is found in Figure~\ref{fig:umap_char0}. Attacks that could not be clustered are labeled in the Figure as cluster $-1$. For each cluster, the archetype attack domain is noted.

\begin{figure}
\centering
\begin{minipage}{0.45\textwidth}
\includegraphics[width=0.95\textwidth]{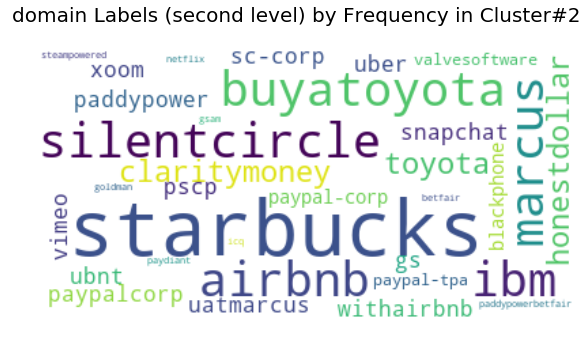}
\caption{Sample Cluster derived from the char0 distributions. This cluster contains a distinct set of domains owned by AirBnB. These attacks have low overlap with the DNS enumeration list.}
\label{fig:airbnb1}
\end{minipage}\hfill
    \begin{minipage}{0.45\textwidth}
    \includegraphics[width=0.95\textwidth]{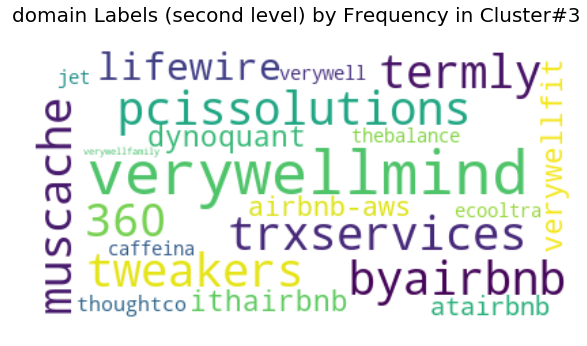}
\caption{Sample Cluster derived from the char0 distributions. This cluster contains a distinct set of domains owned by AirBnB. These attacks have very high overlap with the DNS enumeration list.}
\label{fig:airbnb2}
\end{minipage}
\end{figure}

These clusters also picked up sets of \emph{related} domains, which again is consistent with a single attack generator. As an example, seen in Figure \ref{fig:airbnb2}, one cluster includes numerous domains owned by the company AirBnB, e.g., \texttt{atairbnb.com} and \texttt{withairbnb.com}, but also includes the AirBnB content-delivery-network (CDN), \texttt{muscache.com}. This same cluster contains numerous domains owned by the media company DotDash\footnote{Formerly about.com, DotDash domains include thoughtco.com and lifewire.com, among many others.}. Another cluster is dominated by a combination of Pharmaceutical Industry domains and domains registered in Iceland. Each of these clusters represent attacks over numerous days within the approximately sixty days represented by the data sample. Two clusters contained domains owned by AirBnB, but that did not overlap; see Figures \ref{fig:airbnb1} and \ref{fig:airbnb2}.

There are a few obvious downsides to this approach of feature engineering. First, we made an assumption that while the volume of data we observe may vary widely between attacks, our collection perspective relative to the attack generators is unchanged, and therefore the character distributions from the same attack generator will be relatively consistent over time. This is a useful but flawed assumption that could lead us to separate attacks that are created by the same generator. Second, we treat each set of distributions as derived from a single generator. If multiple generators are used during an attack, each generator will contribute to a mixed character distribution. The expected result is that, in that case, we observe more clusters than attack generators that truly exist.

Considering character distributions at-scale confirmed our suspicion that most attacks in our dataset do not use a pseudo-random prefix generator. In nearly 450 attacks, only eight of the attacks had distributions similar to a uniform random distribution. Even in those cases, it wasn't apparent that the prefixes were pseudo-randomly generated. Attacks on the domain \texttt{nspk.ru} proved to be constructed from incrementally adding sorted characters. For example, \texttt{a.nspk.ru, b.nspk.ru,...aa.nspk.ru, ab.nspk.ru}, etc. So, while the distribution was the same as a pseudo-random generator, the attack was not. An attack on the domain \texttt{verimi.de} observed on July 6, 2018 exhibited a nearly uniform random distribution and was used for a archetype attack for clustering.

Having identified promising clusters, the most central attack within each of these clusters is determined by the shortest distance to each other attack within the cluster. The associated distribution for this central attack is the representative for the cluster and the attack is the archetype for that cluster.  In the final clustering, one feature is generated for each of these archetype representatives: the Jensen-Shannon distance between an attack and the archetype. We also included the distance to the \texttt{verimi.de} attack as a uniform random representative. The distance between cluster centers is shown in Appendix Figure \ref{fig:center-dist}.

\subsection{DNS Enumeration}
\label{sec:dns_enum}

During exploratory data analysis, there was a surprising combination of relatively common labels and seemingly unrelated and uncommon labels within the same attack. We isolated long labels, greater than twenty characters, which were present in multiple attacks. These included a strange mix of dictionary-type labels, composite terms, and multiple languages, such as
\begin{itemize}
    \item \texttt{mobilebusinessapplicationdevelopment},
    \item \texttt{007dapotianmuweijiaomenduchang}, and
    \item \texttt{caracepatmembuatwebsitegratis}
\end{itemize}
Surprisingly, a Google search for one of these terms led to a GitHub repository~\cite{github_dns} containing a list of over 420,000 terms culled from DNS, a so-called \textbf{DNS enumeration list}. This kind of list is generated by cyber actors to assist in reconnaissance and formulating attacks, but the use has not been previously reported in Slow Drip attacks. 

\begin{figure}
    \centering
    \includegraphics[width=\linewidth]{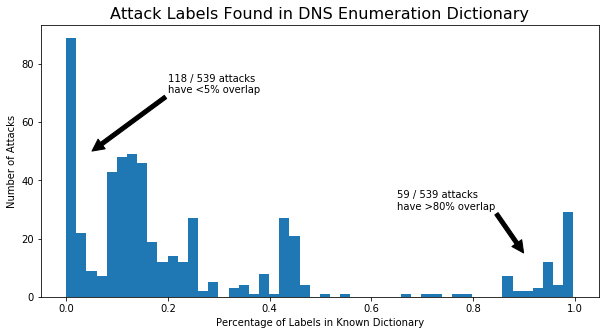}
    \caption{Percentage of labels per attack that overlap with GitHub DNS enumeration list.}
    \label{fig:dns_overlap}
\end{figure}

We hypothesized one of the attack generators utilized a DNS enumeration list as a dictionary, creating FQDNs during an attack by picking words from the dictionary and continually appending them as labels. For each attack the intersection of its labels and the GitHub list was computed. Over 15\% of them drew 80\% or more of their labels from this dictionary, all but confirming that some kind of DNS enumeration list was used as a dictionary for an attack generator, given the volume of unique labels found in attacks. A significant number of attacks contained 30-60\% overlap, as seen in Figure~\ref{fig:dns_overlap}, which could be consistent with either multiple attack generators in use for those attacks or a larger underlying dictionary. 

Manual inspection of the domains found in each grouping showed correlation with the clustering based on character distributions. In particular, the two clusters generated by the Jensen-Shannon distance between character distributions containing AirBnB domains correlated to a high and low inclusion of the DNS enumeration terms in their labels, as noted in Figures \ref{fig:airbnb1} and \ref{fig:airbnb2}.  

The overlap with the DNS enumeration dictionary is not notable in any of the attacks found in data from open resolvers. It's possible that actors using a DNS enumeration dictionary are not leveraging open resolvers in their attacks. 

\subsection{Time Series Features}
\label{sec:timeseries}

We found during the exploratory data analysis phase that when considered as strings in time order, the FQDNs used in an attack were often in \textbf{lexicographical order}, that is, a query for \texttt{aa-dev.airbnb.com} would be transmitted prior to one for \texttt{ab-dev.airbnb.com}. However, capturing this observation into a numerically accurate feature is difficult. For example, an attack that is generated by drawing prefixes from a sorted dictionary and create traffic from multiple locations will be observed as an interleaved set when aggregated in time. The more complex the generator and the more sources of traffic, the more obscured the original ordering becomes. A number of factors could influence ordering, in particular:
\begin{itemize}
    \item Observations are not directly from the attack generators and may contain the interleaving of the attack traffic of several devices.
    \item Internet routing may effect the timing of events at sensors and result in observations that are out of order from their original creation.
    \item The sensors themselves are often part of load balanced systems which further diffuse arriving traffic.
    \item Additional queries may be created by Internet appliances along the routing path, adding noise to the data.
    \item A large number of events occur every second, the smallest granularity of our observation period, and we are unable to determine the order in which those are generated. 
    \item It is extremely unlikely all events are observed.
\end{itemize}  

\begin{figure}
    \centering
    \begin{minipage}{0.45\textwidth}
    \includegraphics[width=0.95\textwidth]{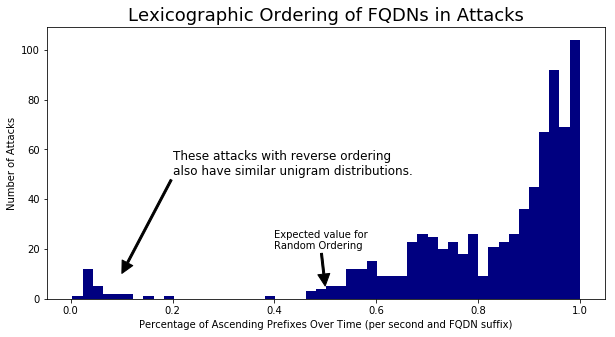}
    \caption{An approximation of the lexicographic ordering of events within attacks estimated by considering a single FQDN per suffix per second.}
    \label{fig:lexo}
    \end{minipage}\hfill
    \begin{minipage}{0.45\textwidth}
    \includegraphics[width=0.95\textwidth]{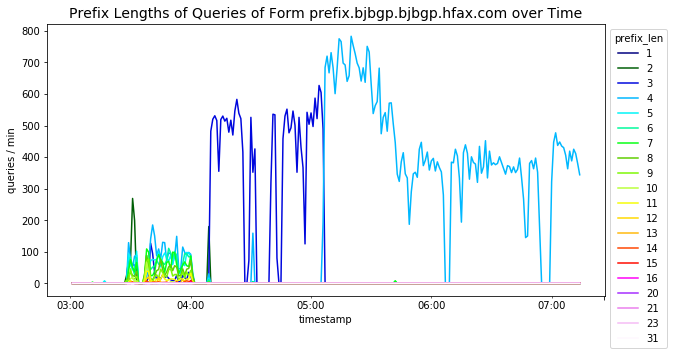}
    \caption{Prefix lengths observed over time for an attack against hfax.com. For a fixed suffix, the prefix lengths are initially quite variable, but transition to length three, and then length four, for extensive time periods.}
    \label{fig:hfax}
    \end{minipage}
\end{figure}

To address these challenges, we \textbf{approximate the lexicographic ordering} by drawing a single event for each \textbf{suffix}\footnote{Recall, the suffix of an FQDN is found by removing the prefix, $label_n$.} per second, creating a time series for each suffix. The suffix is fixed because exploratory analysis led to the hypothesis that subdomains may be created incrementally by increasing the number of labels in the FQDN. Given a fixed suffix, and a sample for each second, we calculate the percentage of these that are in order. For each attack, the average of this percentage over all suffixes is an approximate measure of the lexigraphic ordering. We represent this as a ratio of events in the interval $[0,1]$, as shown in Figure \ref{fig:lexo}. Somewhat surprisingly, almost no attacks contain randomly ordered prefixes. A majority of attacks have strong tendency toward lexicographic ordering, while twenty-seven of the attacks demonstrate reverse ordering.

Another feature drawn from the exploratory data analysis was the length of each attack. While most attacks were relatively short, lasting under fifteen minutes, several attacks lasted hours and some nearly a full day. Obtaining an accurate measure of the attack length was hindered by noise in the traffic and skew in the distribution of events per minute for each domain. Popular domains are likely to have a constant stream of unresolvable queries, due to factors such as user typos and the use of content delivery services. Measuring an attack by the number of minutes for which there were more unresolvable queries than the mean, a common benchmark statistic, led to inaccurate results given these factors. We found the number of minutes in which attack events exceeded the median number of non-existent subdomains (NXDOMAIN events) measured over the full day more accurate. This value was used as the \textbf{attack length} feature in clustering.

Time series analysis indicated there was a reasonable likelihood that multiple attack generators were used in some attacks. The number of labels and variety of prefix lengths observed per second were illuminating. In a small subset of attacks there were distinct periods of variable behavior. An example of this shown in Figure \ref{fig:hfax}, in which a period of widely varying prefix lengths is followed by long periods, with far greater activity, containing a single prefix length and a smooth transition between lengths. Quantifying these behaviors over all attacks, however, would be extremely difficult, and these observations were not used in the final clustering algorithms.

\subsection{Qtype Features}

Another striking observation from exploratory data analysis was the use of multiple DNS query types (\emph{qtype}) during an attack. This has not been reported previously. As mentioned in Section \ref{sec:dns}, while DNS is most recognized for its translation of domain names to IP addresses, it can deliver a multitude of other types of information. The type of data sought is included in the query. For example, a query for type "MX" would request the mail server domain, rather than the IP address of a domain. While IPv4 address requests, called "A" records, are by far the most common request seen in DNS traffic, the protocol allows for 65536 different data types. Other commonly seen query types includes the IPv6 address, type "AAAA", and the authoritative name server, type "NS". Slow Drip attacks are known to request IPv4 (A) records~\cite{ExploderBot}. 

Many attacks included queries for less common query types, and these occurred in distinct combinations. Six record types were often simultaneously utilized, though some variations existed. These six record types were query types: 1 (IPv4 address), 5 (CNAME), 12 (PTR), 16 (TXT), 28 (IPv6 address), and 33 (SRV). Of these, PTR, CNAME, and SRV are the most unusual. We generated \textbf{qtype features} for each of these six by computing the percentage of events in an attack of each type.

We also observed large queries for a handful of other unexpected qtypes. A comparison of attacks using multiple qtypes to those with only "A" record queries is shown in Appendix Figure \ref{fig:qtypes}. Both the combination of query types and the subdomains found within each type were rare within normal DNS traffic. From an attacker's perspective, this variety of query types serves to further diffuse the attack and ensure that the majority of requests are forwarded to the authoritative name server.

\section{Final Cluster Analysis}\label{sec:clusters}

An unsupervised learning technique, \emph{Hierarchichal Density-based spatial clustering of applications with noise (HDBSCAN)} \cite{HDBSCAN}, an extension of DBSCAN~\cite{DBSCAN}, was used to cluster attacks.\footnote{A minimum cluster size of ten was used.} None of the features considers either the date of the attack or the attacked domain itself, which are the two components used to uniquely label an attack. Thus the clusters describe attacks that have generative features in common and represent attacks that are likely created by the same attack generator. This clustering was performed on the set of 435 attacks found from June-August 2018 from the larger data source and independently applied to the attacks found in the open resolver set for source comparison. Finally, the same features were used to cluster data from January 2019 to ensure continued relevance of the features and clusters. 

\subsection{Feature Set}

\begin{figure}
    \centering
    \includegraphics[width=\linewidth]{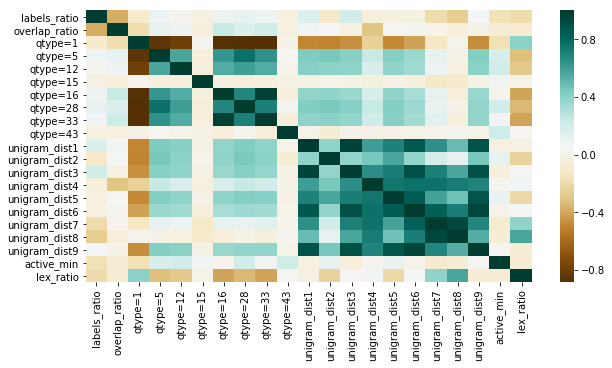}
    \caption{Correlation of twenty features used for clustering.}
    \label{fig:feature_corr}
\end{figure}

We used a set of twenty features based on the initial analysis in Section~\ref{sec:agenda}. These considered the uniqueness of the domain labels, their overlap with the DNS enumeration dictionary\footnote{Here we mean the list found on GitHub discussed earlier.}, the distance between unigram character distributions and nine established distributions (including one uniform random distribution), the length of the attack, the variety of query types (qtype), and an estimate of the lexicographic ordering of the queries. The features are detailed in Table~\ref{tab:features}. The correlation of the features is shown in Figure \ref{fig:feature_corr}. We can also use Uniform Manifold Approximation and Projection (UMAP)~\cite{UMAP} \cite{UMAP_software} to reduce the features into two-dimensional space for the purpose of visualization. This projection allows us to visualize how similar attacks are with one another, as seen in Figure~\ref{fig:umap_features}.

\begin{table*}[]
    \centering
    \begin{tabular}{p{0.05\linewidth}p{0.15\linewidth}p{0.7\linewidth}}
    \hline
        Count & Name & Description \\
        \hline
        \hline
        1 & $label\_ratio$ & percentage of unique labels in attack \\
        1 & $overlap\_ratio$ & percentage of label overlap with DNS enumeration dictionary \\
        6 & $qtype=n$ & percentage of query type $n$ records in attack, $n$ in [1, 5, 12, 15, 16, 28, 33, 43] \\
        1 & $active\_min$ & number of minutes where NXDOMAIN events greater than median of all events per minute \\
        1 & $lex\_ratio$ & estimated percentage of lexicographically ascending subdomains \\
        9 & $unigram\_distn$ & J-S Distance from Unigram Char0 Distribution for these attacks as (date, $attack\_domain$): (2018-08-13, \texttt{shave.io}), (2018-08-19, \texttt{informatica.com}), (2018-07-24, \texttt{toyota.com}),
        (2018-07-31, \texttt{ecooltra.com}), (2018-08-07, \texttt{centralnic.com}), 
        (2018-07-12, \texttt{verimi.de}), (2018-08-09, \texttt{norwichpharma.com}),
        (2018-08-02, \texttt{ruckuslbs.com}) \\
        \hline
    \end{tabular}
    \caption{Features Used for Clustering Attacks}
    \label{tab:features}
\end{table*}

\subsection{Results}

\begin{figure}
    \includegraphics[width=\linewidth]{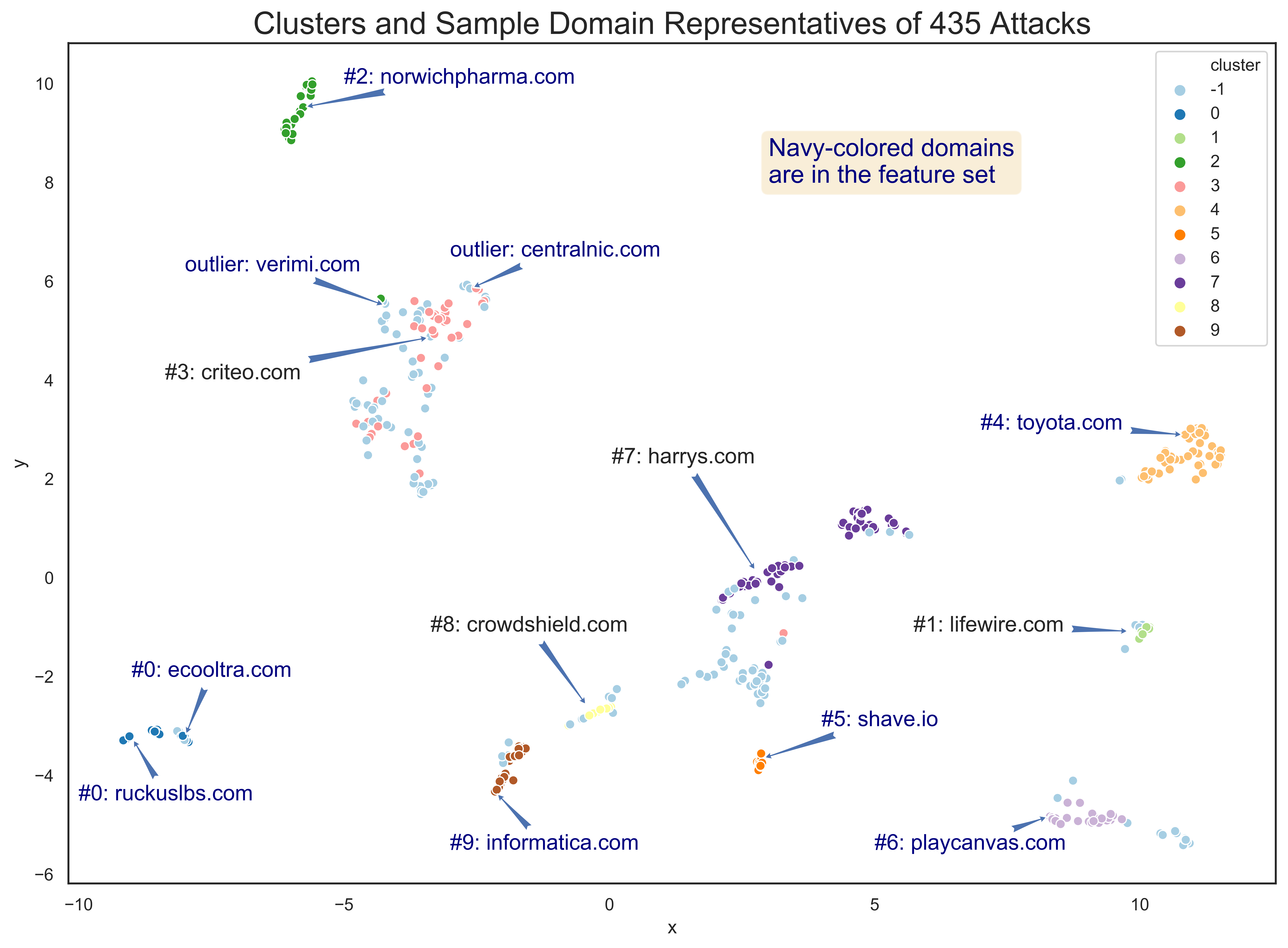}
    \caption{Clusters in a sample of 435 attacks using features from Table \ref{tab:features} and labeled with sample domains. Visualization via UMAP~\cite{UMAP_software} projection.}
    \label{fig:umap_features}
\end{figure}

The clustering revealed nine large groups of attacks, accounting for 279 of the 435 attacks. These clusters and their dominant features are described in Table \ref{tab:clusters}. A visualization of the clusters is found in Figure~\ref{fig:umap_features}.\footnote{This visualization, and other similar ones in this paper, are generated with Uniform Manifold Approximation and Projection (UMAP)~\cite{UMAP}, a means to project high-dimensional data into a smaller space.} Attacks that could not be clustered automatically are in cluster $-1$ indicated as pale blue points. From the plot we can see that although the clustering algorithm was unable to include these points into a specific cluster, in most cases they appear likely to belong to one of the named clusters. Were the attack generators highly variable, we would observed unclustered, or outlier, points scattered across Figure~\ref{fig:umap_features}.

These clusters are similar, but not identical, to those found using only Jensen-Shannon distance between unigram character distributions, as evidenced by two of the unigram attacks falling into the same cluster, and two others as outliers. Each cluster crosses a range of dates, but often contains obviously related attack domains. Several smaller clusters were also apparent, and if we allow clusters as small as five attacks, the clusters represent 293 of the attacks. We suspect that a large number of attacks are created by a single system, but also that two generators are in use at the same time, hindering easy identification.

The clustering of attacks found within open resolver data is also illuminating. If we consider attacks that are observed at \emph{both} multiple data sources, we find that they almost entirely fall within a single cluster, cluster $6$ in Table~\ref{tab:clusters}, represented by \texttt{playcanvas.com}. Broadening the scope to include common attack domains, and not specific attacks, the open resolver attacks primarily fall into two clusters, cluster $6$ and cluster $4$.

\begin{table*}[]
    \centering
        \begin{tabular}{ p{0.03\linewidth}p{0.03\linewidth}p{0.03\linewidth}p{0.35\linewidth}p{0.45\linewidth} }
        \hline
        No. & Size & Days & Sample Attack Domains & Dominant Features \\
        \hline
        \hline
            0 & 20 & 6 & AirBnB-related, Ruckus Wireless & high DNS overlap, five qtypes, lex-ascending, moderate attack lengths  \\
            1 & 11 & 5 & DotDash-related &
            redundant labels, one qtype, high DNS enumeration overlap, short attacks \\
            2 & 28 & 5 & Pharmaceutical Industry and Icelandic &  unique labels, lex-descending, five qtypes, moderate length attacks \\
            3 & 46 & 28 & Indonesian domains, universities, payment sites, Tinder & 
            very lex-ascending, five qtypes \\
            4 & 47 & 19 & Toyota, Starbucks, Paypal  & high label reuse, high DNS overlap, one qtype \\ 
            5 & 10 & 9 & \texttt{dollarshaveclub.com}, \texttt{shave.io} & no DNS enumeration overlap, one qtype, strong unigram pattern, shorter attacks \\
            6 & 27 & 8 & {starting \texttt{sc-}, AirBnB-related, PlayCanvas} & 
            moderately duplicated labels, one qtype, no DNS enumeration overlap, lex-ascending, strong unigram pattern \\
            7 & 55 & 12 & Western Union, Harrys & \textasciitilde50\% label reuse, one qtype, short attacks, high lex-ascending \\
            8 & 11 & 6 & \texttt{crowdshield.com}, \texttt{coupa.com} & one qtype, high lex-ascending, far from unigram models \\
        \hline     
        \end{tabular}
        \caption{Clusters of at least ten attacks within a dataset of 435 attacks selected over 60 days using twenty features in Table~\ref{tab:features}.}
        \label{tab:clusters}
\end{table*}

\subsection{Resilience to Model Drift}

A significant potential Achilles heel for machine learning is known as \emph{model drift}, in which the underlying data landscape changes and a model that previously performed well begins to fail. In the context of Slow Drip attacks, this would be the case for any detection based on the assumption of pseudo-random subdomains, for example, as the threat landscape has evolved. For this reason, models need to be regularly checked for continued relevance to the environment. The original feature engineering and clustering was computed from data for attacks that occurred in June-August 2018. We extracted these same features from nine days of attacks in January 2019 to determine whether they remained valid. During this time there were 106 attacks detected on 83 domains. Far more attacks in this new data set overlapped with the DNS enumeration list~\cite{github_dns}, with over 25\% of the attacks drawing more than 80\% of their labels from the list. 

To evaluate the relevance of the features in Table~\ref{tab:features} to the new attacks, HDBSCAN was used to cluster the new attacks, as well as the union of the two datasets. When considered alone, 93 of the new 106 attacks fall into two clusters, indicating that the features remained relevant for clustering the new data. When combined with the earlier data set, 64 fell into previously defined clusters, indicating stability in the clusters over six months. Most of the attacks in the January 2019 data fall into cluster $0$ in Table~\ref{tab:clusters}, while the others fall into cluster $6$ and cluster $4$. 

\section{Conclusions}\label{sec:conclusions}

The results of this research and development support both threat analysts and data scientists. An anomaly based statistical classifier provides a means to detect these attacks with high reliability in both very large-scale and moderate-scale data environments.  Feature engineering led to the discovery of notable characteristics of several generators, including the use of DNS enumeration dictionaries and unusual query type combinations, that can help analysts recognize patterns in putative attacks. It further confirmed their concerns that the threat landscape had changed, making the attacks harder to recognize. Unsupervised machine learning identified nine major clusters and the associated important features for each cluster. Moreover, the features remained relevant after six months.  

For data scientists working in Cyber Security, the features described here are markedly different from those commonly described in the application of machine learning to problems leveraging DNS data. The approaches provide alternate tools that may apply to a wide range of problems. In particular, while text-based features are frequently used in the literature, we have not seen the similarity of distributions included. Similarly, the use of query types and the lexicographical ordering were not seen elsewhere. 

In the remainder of this Section we summarize the findings and our conclusions, organized by the associated threat intelligence questions.

\subsection*{How is attack traffic generated?}

The current threat landscape does not resemble previously described attacks, and, in particular, pseudo-random subdomains are not prevalent. Only eight of the attacks contained nearly uniformly random character distributions. Even those attacks didn't all show signs of pseudo-random generators. There was no evidence of the particular pseudo-random generator used in the ExploderBot system.

In contrast, many of the attacks appear to be dictionary generated. Over 15\% of the attacks appear to be generated through the use of a DNS enumeration dictionary. That generator appears to traverse the list in lexicographic order and continually append new labels. While we don't have the underlying dictionary used, the list includes almost every term of a 420,000 word list located in a GitHub repository~\cite{github_dns}. Other attacks similarly had significant overlap with Unix dictionaries. 

When considered as events in time, over 80\% of the attacks consisted of FQDNs that were ordered lexicographically, or alphabetically, rather than randomly. However, we also discovered one cluster of twenty-eight attacks that demonstrated descending order, shown in Table ~\ref{tab:clusters}. These findings are consistent with the use of sorted dictionaries as a source for generators.

Further, attackers have broadened from the original attack scenario to incorporate multiple query types. This approach allows them to use the same set of FQDNs with variable query types, assuring the queries are forwarded to the authoritative server while minimizing the dictionary they might maintain. In doing so, they created a signature in the choice of qtypes used within an attack.  

\subsection*{How many systems are active?}

Clustering results indicate that attack generators are limited to a relatively small number of algorithms. Multiple clusters shown in Table \ref{tab:clusters} indicate the use of several query types, moreover they contain a combination of uncommon query types. We suspect that there are commonalities in these generators, even where their subdomain features are distinct, based on this query type pattern. Combining the clustering results with subject matter expertise, there are likely four to eight attack generators in use that are able to create global-scale attacks on repeated days. The projection of the twenty features into two-dimensional space, Figure~\ref{fig:umap_features} is consistent with this conclusion. 

Some attacks appear to be a composite of multiple generators. Figures \ref{fig:hfax} and \ref{fig:harrys} show two examples where we see abrupt changes in statistical features of the attack. In addition some of the attacks appear to be a composite of a DNS enumeration dictionary generator and another dictionary source. In addition, rapid changes in the characteristics of the FQDNs, such as the prefix changes seen in Figure~\ref{fig:prefix_ts} were fairly common and often overlapping. Further, the projection of clusters in Figure~\ref{fig:umap_features} shows that many of the attacks that were not clusters lie in between those that were, indicating they might be a mix of the two cluster generators.
 
\section{Future Work}

There are numerous avenues for further research into this area the results of which can help us understand cyber actors better and may lead to techniques that can be applied to a broader set of problems. In our research, we did not study the victims, for example, which may further refine our understanding of both the generators and the actors operating these attacks. One could take a number of graph approaches to these problems, including a study of how the labels form tightly connected clusters among attacks. Similarly, one could study the qtype distribution across attacks as a bipartite graph. The projection of the attacks into two dimensions as shown in Figure~\ref{fig:umap_char0} raises additional questions about the underlying cause for linear patterns in the projected points.

The incorporation of open resolvers to diffuse an attack as part of an actor's Techniques, Tactics, and Procedures (TTP) can help fingerprint their activity. While we looked at global-scale attacks also observed at open resolvers, an in-depth study of focused on attacks observed in open resolvers could shed light on the breadth of actor's incorporating this tactic in their attacks. 

Finally, techniques to separate multiple generators used during a single attack could help refine our understanding of both the generator techniques and the actors. Our analysis appeared to indicate multiple generators might be used in sequence, demonstrated by abrupt changes in the generator over time, as well as concurrently, demonstrated by a mix of dictionary and crafted labels. 

We have not seen distribution distance measures used as a clustering feature in the manner used here. The approach of comparing character distributions, particularly when combined with reference distributions, scales well and can be applied to other problems where we have large samples of strings generated through different mechanisms. One alternate application is in the application to malware utilizing Domain Generation Algorithms (DGA).

The clusters, and respective data, found in Table \ref{tab:clusters} can be used to create classifiers for detected attacks. This would allow attacks to be labeled and associated with other attacks, while reserving analytic resources to attacks that are not able to classified automatically.  

\section*{Acknowledgments}

I'd like to thank the Infoblox Threat Analysts and Cyber Intelligence Director, Sean Tierney, for their insights into the Slow Drip problem from a threat analyst perspective and help focusing this work for security applications. Nathan Toporek, an Infoblox Threat Analyst, also helped in early feature engineering. Mike Last provided statistical advice that helped guide the direction of this research. Laura da Rocha, Chris Heckman, Mike Last, and Cameron Switzer all proved dedicated reviewers, making the paper clearer and more relevant. This work heavily relied on Open Source Software and Python libraries, without which research of this complexity would be extraordinarily difficult to accomplish. 

\newpage
\appendix
\section{APPENDIX: Supporting Visualizations and Additional Features}

We've included visualizations and further details in the Appendix that might appeal to certain readers interested in the feature engineering performed. In addition, this appendix includes a number of features we evaluated in the course of our research but did not use in the final clustering. 

\subsection{Unigram Distributions as Features}

\begin{figure}[h]
\centering
\begin{minipage}{0.45\textwidth}
\centering
\includegraphics[width=0.95\textwidth]{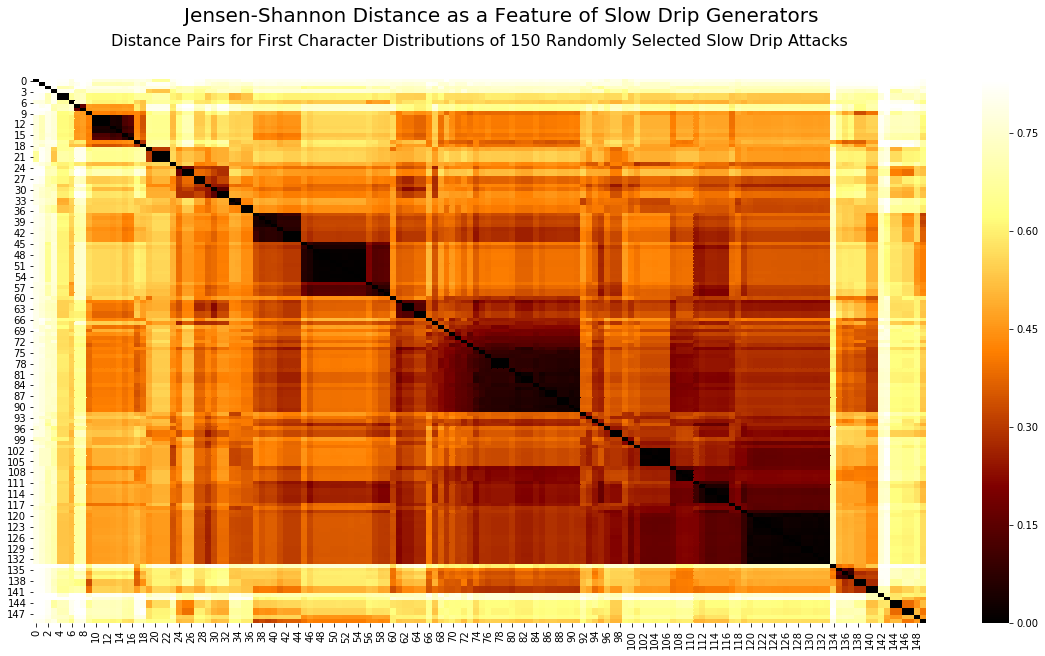}
\caption{Pairwise Jensen-Shannon distances between char0 distributions of prefixes in 150 randomly selected attacks, sorted to clustering of similar distributions.}
\label{fig:JSD}
\end{minipage}\hfill
\begin{minipage}{0.45\textwidth}
\centering
\includegraphics[width=0.95\textwidth]{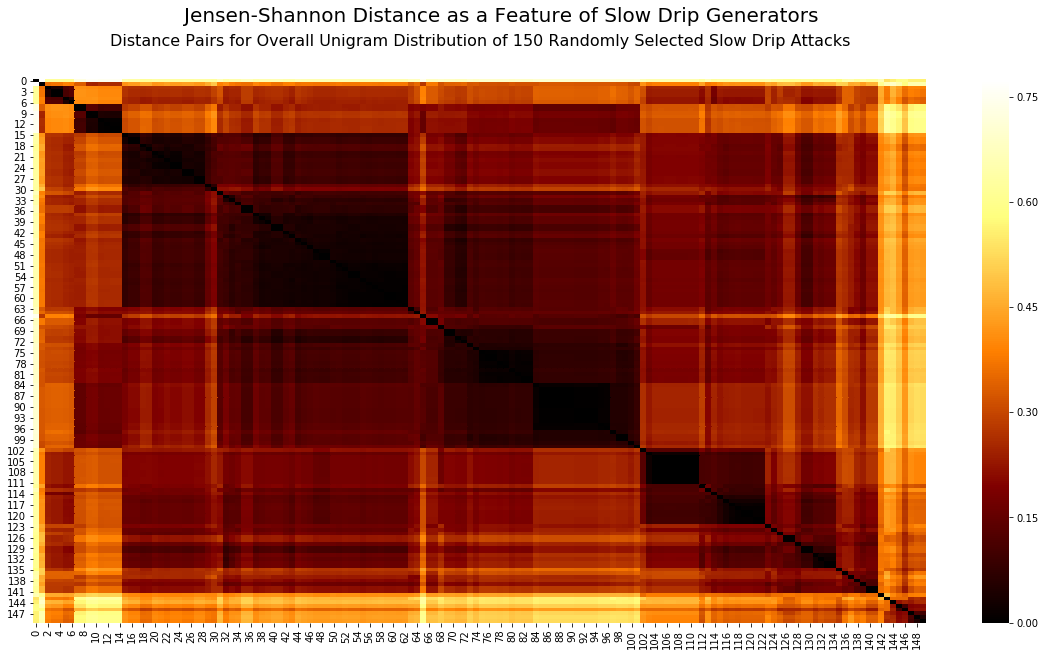}
\caption{Relative pairwise Jensen-Shannon distance between overall unigram distributions of prefixes in 150 randomly selected attacks. While clusters exist, they are far less distinct than those found using the char0 distributions.}
\label{fig:JSDoverall}
\end{minipage}
\end{figure}

Visualizing the similarity, or J-S distance, between distributions, with a heat map, confirms that this distance measure can be used to separate attacks. The distance between attacks is the J-S distance between the respective character distributions for each attack, and is in the range $[0,1]$. Figure \ref{fig:JSD} shows a heat map of pairwise distances, with smaller distances represented by darker pixels. In this random sample of 150 attacks, subsets of similar attacks are seen as dark groupings along the diagonal.\footnote{This visualization also uses agglomerative clustering of the results to sort the attacks.} Attacks that have dissimilar char0 distributions will have light-colored pixels. Considering this, in Figure~\ref{fig:JSD} we see a few outlier attacks in the top left, some smaller groups of similar attacks, and a larger set that shows core similarity with broader distribution.

If we compare the char0 distribution similarities with the overall distribution similarities, we see a dramatic difference. The distance matrix for overall unigram distributions of the same set of 150 attacks is illustrated in Figure~\ref{fig:JSDoverall}. Here we observe much less variance in the distributions. 

Recall from Section~\ref{sec:unigram} that the attacks were clustered via their char0 distributions and eight clusters identified, the centers of which were considered \emph{archetype} distributions. The distances between these archetype distributions is shown below in Figure~\ref{fig:center-dist}. This allows the reader to see that the clusters are reasonably separated, though a few are closer and are possibly related distributions. As these distributions are later used in clustering all of the attacks, we want to observe some separation in the data. Our distributions include a nearly uniform random one, generated by an attack on \texttt{verimini.de}, that is furthest from all other distributions. 

\begin{figure}
    \centering
    \includegraphics[width=\linewidth]{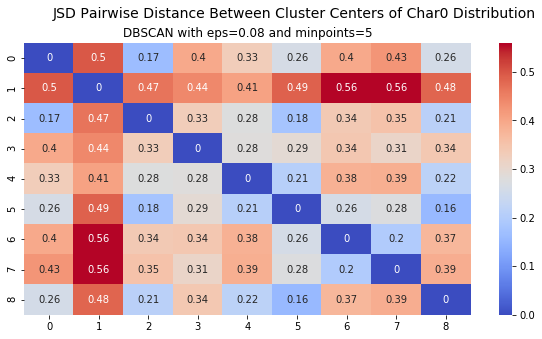}
    \caption{The Jensen-Shannon distances between char0 cluster centers indicates good separation between the clusters.}
    \label{fig:center-dist}
\end{figure}

\subsection{Query Types as Features}

\begin{figure}
    \centering
    \begin{minipage}{0.45\textwidth}
    \centering
        \includegraphics[width=0.95\textwidth]{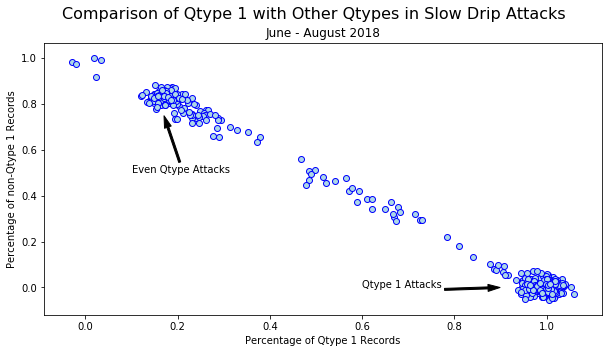}
    \caption{The relative type distribution between type 1 ("A") record queries and other types over 444 attacks, including small jitter to the x- and y- coordinates to visually separate attack points.}
    \label{fig:qtypes}
    \end{minipage}\hfill
    \begin{minipage}{0.45\textwidth}
    \centering
    \includegraphics[width=0.95\textwidth]{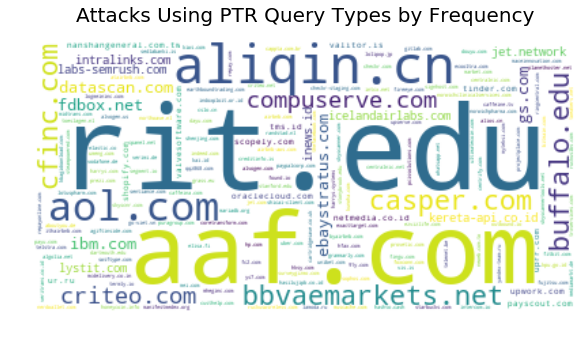}
\caption{Domains attacked using PTR queries sized by the number of PTR records observed.}
\label{fig:ptr}
    \end{minipage}
\end{figure}

Most attacks contained either a single query type, qtype 1, or a specific, surprising combination of five query types. In Figure~\ref{fig:qtypes} a scatter plot shows the proportion of qtype $1$ queries in attacks. Notice that there are two large groupings, but also a small set of attacks with a wide degree of variance. 

Another observation of the exploratory data analysis, was the transition in some attacks between the use of several query types to a single query type over time. This phenomena, shown in Figure \ref{fig:harrys}, serves to hinder using the query type alone as a strong feature. In this case, six query types are used simultaneously for part of the attack, five in even proportions, and then a transition occurs to a single query type. This type of phenomena occurs in a small, but notable, portion of the attacks. 

Nearly half of the attacks contained a substantial number of qtype 12 (PTR) records requests. This is particularly surprising, as PTR records are primarily used to locate a hostname from an IP address in a specially formatted FQDN. An example of the domains attacked using PTR records are shown in Figure \ref{fig:ptr}. Notably, none of the AirBnB-related attacks contain PTR records, whereas attacks against universities, e.g., \texttt{rit.edu}, all contain PTR records. 

\subsection{DNS Enumeration as a Feature}

\begin{figure}
    \centering
    \begin{minipage}{0.45\textwidth}
    \centering
    \includegraphics[width=0.95\textwidth]{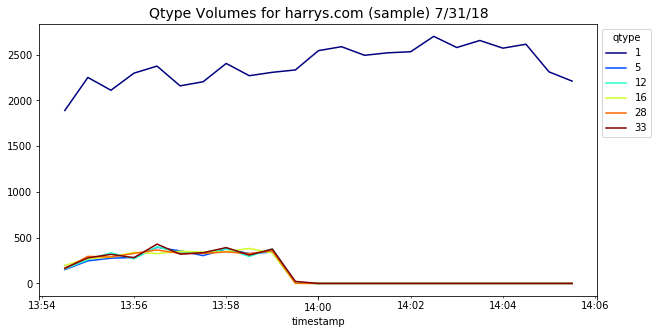}
    \caption{An example of the distribution of query types transitioning over time during an attack from six types to a single type.}
    \label{fig:harrys}
    \end{minipage}\hfill
    \begin{minipage}{0.45\textwidth}
    \centering
\includegraphics[width=0.95\textwidth]{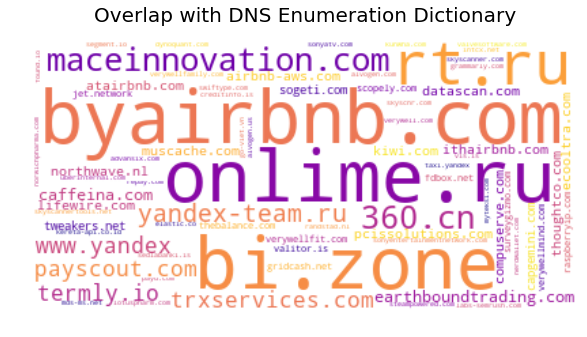}
\caption{Attacked domains, sized by their overlap with the percentage of prefixes found in the DNS enumeration dictionary.}
\label{fig:dns_enum}
    \end{minipage}
\end{figure}

Those attacks containing a very large portion of GitHub DNS Enumeration dictionary~\cite{github_dns}, contained domains that were notably different than seen in other features. While \texttt{byairbnb.com} is present, most of the domains are less recognizable, and quite a few of the larger attacks are against Russian or Chinese domains. Figure~\ref{fig:dns_enum} shows a word cloud containing the attacked domains sized by the percentage of overlap between attack prefixes and the DNS enumeration dictionary.  

While the overlap with this specific list was present in a substantial number of domains, the true dictionary is unknown. We attempted to 'bootstrap' from the attack labels to the underlying dictionary by taking the union of all labels from attacks with greater than an 80\% overlap to the GitHub list. This did increase the dictionary size, but not substantially. More importantly, it didn't impact the overall distribution of the overlap statistic.   

\subsection{Attack Length as a Feature}

\begin{figure}[ht] 
    \centering
    \begin{minipage}{0.45\textwidth}
        \includegraphics[width=0.95\textwidth]{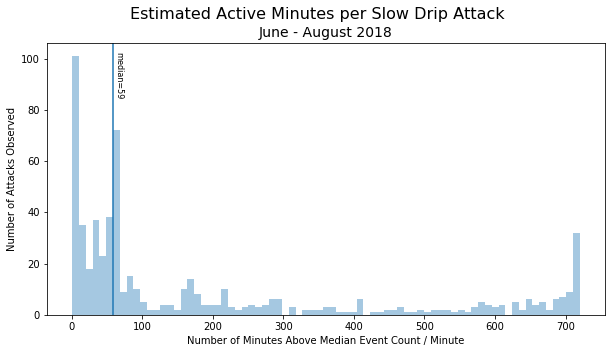}
    \caption{Distribution of attack lengths as measured by the number of minutes in which nxdomain events exceeded the median for the attack domain. }
    \label{fig:length}
    \end{minipage}\hfill
    \begin{minipage}{0.45\textwidth}
    \centering
    \includegraphics[width=\linewidth]{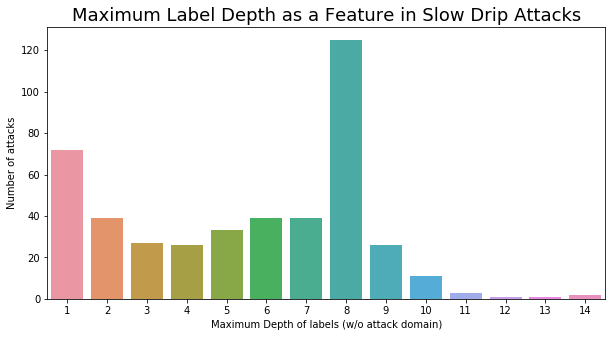}
    \caption{Maximum depth of labels observed during attacks.}
    \label{fig:depth}
    \end{minipage}
\end{figure}

As seen in Figure~\ref{fig:length}, most attacks are quite short. This is consistent with other studies of DDoS attacks writ large~\cite{JonkerDDoS}. Accurately estimating the length of an attack using a single computation across domains is difficult due to the variance in baseline traffic for each domain which creates noise. The common approach of using a mean will be inaccurate due to the skew within the distribution of events per minute for various types of domains. We evaluated multiple measures, and ultimately used the number of minutes for which NXDOMAIN responses for a domain were above the median level for that domain during the day. This measure tends to underestimate attack lengths for rare domains, e.g., \texttt{hfax.com}, and overestimate lengths of popularly domains, e.g., \texttt{airbnb.com}, but creates a reasonable relative measure.  The vast majority of long-lived attacks are unpopular domains, as measured by global DNS requests over time. 

\subsection{Other Potential Features}

We evaluated a wide range of features, many of which had promising characteristics, before settling on those used in this paper. In this Section, we provide an overview of some of the other features we considered and that might be relevant for understanding actors and attack generators. 

\subsubsection{Open Resolvers Usage}

While we did not utilize the presence of an attack as an explicit feature in our clustering, the use of open resolvers by an attacker to diffuse their packets is a tactic that may serve to fingerprint the attack generator. Our results found that the attacks observed at an open resolver were all within one cluster in Table~\ref{tab:clusters}.  

For attacks created through consumer-based botnets, there is little advantage to using open resolvers and there are evident risks.\footnote{By consumer-based botnets, we mean botnets of compromised devices found in a home consumer environment, such as laptops, phones, and home routers, as well as Internet of Things devices like alarm and lighting systems.} In many commercial environments, DNS requests to external IP addresses will be blocked, thereby thwarting an attack emanating from these networks. Additionally, the space of open resolvers changes and the bot members need to be regularly updated. A simpler solution is to have bot members transmit requests through their normal recursive resolver. On the other hand, attacks generated from an actor-controlled infrastructure, such as cloud services provider, benefit from scattering the initial DNS requests across open resolvers without much risk. 

Our focus in this research was on global-scale attacks, from which it is relatively difficult to discern how widely open resolvers are utilized given the natural transit of traffic through the Internet; most of our observations occur at a point where the connection with the initial request to an open resolver would be lost. A more in-depth look at all attacks found at various open resolvers may shed further light on generators. 

\subsubsection{Label Depth}

While the Slow Drip is typically characterized as a single random label on a fixed attack domain, as shown in \ref{eq:slowdrip}, we found that current attack systems frequently used more labels. The \emph{depth} of an FQDN is the number of labels it contains. If we remove the attack domain from each FQDN within an attack, we can consider how many labels, in terms of depth, were used in the attack. Instead of a single random label, we find the most attacks contain more depth, and that eight labels is particularly common; see Figure \ref{fig:depth}. The variance seen here indicates that this feature can be used to separate some of the attacks, however overall it is a weak feature. We also considered averaging the unique levels per minute to incorporate time, and alternate descriptive statistics such as the variance over time. As seen later in Figure~\ref{fig:labels_ts}, however, this simple descriptive statistic can obscure strong time series features.

\subsubsection{Prefix Lengths}

Another natural set of features to investigate surround the label or prefix lengths. The distribution of these lengths appears to vary with some consistency across attacks. We examined various descriptive statistics for the prefix lengths, including the standard deviation and co-variance per attack, but ultimately did not use these in our clustering.  

\subsubsection{Time Series Features}

\begin{figure}
    \centering
    \begin{minipage}{0.45\textwidth}
    \centering
    \includegraphics[width=0.95\textwidth]{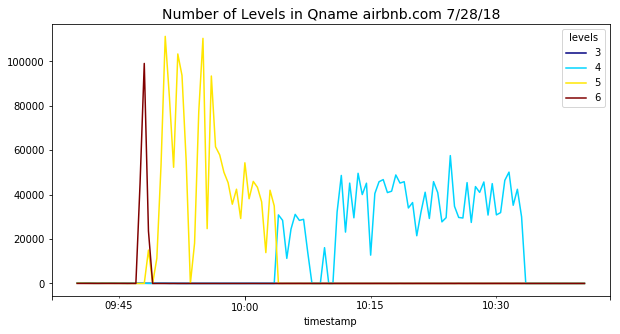}
    \caption{The number of labels over time for one attack on \texttt{airbnb.com}. }
    \label{fig:labels_ts}
    \end{minipage}\hfill
    \begin{minipage}{0.45\textwidth}
    \centering
      \includegraphics[width=0.95\textwidth]{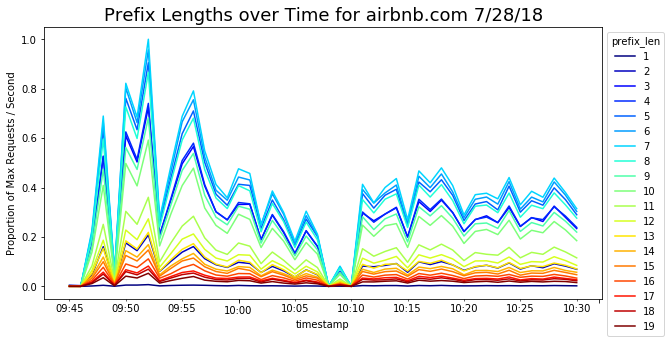}
    \caption{Prefix lengths over time during one attack on \texttt{airbnb.com}. The proportion of lengths remains relatively constant although the volume changes.}
    \label{fig:prefix_ts}  
    \end{minipage}
\end{figure}

During our data exploration and feature engineering, time series analysis again and again revealed characteristics in the data not otherwise visible. Some of these features are seen in the main paper, including the lexicographic ordering in attacks. There were a number of time-related features that we did not capture in our clustering, but found particularly compelling.  Some generators appear to use a fixed number of labels at any time, which changes over the attack, as seen in Figure~\ref{fig:labels_ts}, presumably from building attack queries by continually appending, or removing, labels. In this particular example, the number of labels descends in time, but in many other examples it is seen to ascend. This would be consistent with the use of a dictionary, which is used to select a label, and creating queries by continually appending labels. 

In the same attack shown in Figure~\ref{fig:labels_ts} on \texttt{airbnb.com}, we find that the relative distribution of prefix lengths remains relatively constant over time, even as the number of labels changes. In Figure~\ref{fig:prefix_ts} we see a condensed view of this effect; for space, we have limited the prefix length to twenty. In reality, the longest prefix is fifty-five characters, the maximum length permissible. This is consistent with with the use of an underlying dictionary.

Features considered to incorporate time series elements of the attacks included averaging descriptive statistics such as minimum or maximum of different components, e.g., labels or prefix lengths, per minute over an attack, as well as variance and co-variance measures. While promising, these features proved more difficult to assess and scale given the volume of data and were not used in the final clustering. Given the strong time elements in DDoS attacks, improved approaches to leveraging time series features at scale would likely prove valuable to understanding the malware and actors. 

\newpage

\printbibliography

\end{document}